\documentclass[sigconf,authorversion,nonacm]{acmart}
\usepackage{popets}
\setcopyright{none}

\usepackage{titlesec}
\usepackage{xcolor}
\usepackage{framed}
\usepackage{amsmath}
\usepackage{subcaption}
\usepackage{tabularx} 
\usepackage{tabularray}
\usepackage{multirow}
 \usepackage{colortbl} 
 \usepackage{multicol}
\usepackage{tikz}
\usepackage{array}

\usepackage[most]{tcolorbox}

\newlength{\leftbarwidth}
\newlength{\leftbarsep}
\renewenvironment{leftbar}{%
    \MakeFramed {\advance \hsize -\width \FrameRestore }%
}{%
    \endMakeFramed
}

\begin{document}


\title{SoK: Evaluating 5G Protocols Against Legacy and Emerging Privacy and Security Attacks}

\author{Stavros Eleftherakis}
\email{stavros.eleftherakis@imdea.org}
\affiliation{%
  \institution{Imdea Networks Institite}
  \institution{Universidad Carlos III de Madrid }
  \city{Madrid}
  \country{Spain}
}

\author{Domenico Giustiniano}
\email{domenico.giustiniano@imdea.org}
\affiliation{%
  \institution{Imdea Networks Institite}
  \city{Madrid}
  \country{Spain}
}

\author{Nicolas Kourtellis}
\email{nicolas.kourtellis@telefonica.com}
\affiliation{%
  \institution{Telefonica Research}
  \city{Barcelona}
  \country{Spain}
}



\begin{abstract}
Ensuring user privacy remains a critical concern within mobile cellular networks, particularly given the proliferation of interconnected devices and services. In fact, a lot of user privacy issues have been raised in 2G, 3G, 4G/LTE networks. Recognizing this general concern, 3GPP has prioritized addressing these issues in the development of 5G, implementing numerous modifications to enhance user privacy since 5G Release 15. In this systematization of knowledge paper, we first provide a framework for studying privacy and security related attacks in cellular networks, setting as privacy objective the User Identity Confidentiality defined in 3GPP standards. Using this framework, we discuss existing privacy and security attacks in pre-5G networks, analyzing the weaknesses that lead to these attacks. Furthermore, we thoroughly study the security characteristics of 5G up to the new Release 19, and examine mitigation mechanisms of 5G to the identified pre-5G attacks. Afterwards, we analyze how recent 5G attacks try to overcome these mitigation mechanisms. Finally, we identify current limitations and open problems in security of 5G, and propose directions for future work. 

\end{abstract}

\keywords{privacy, security, identifiers, adversaries, cellular networks}

\maketitle

\section{Introduction}


In the digital society of today, the widespread use of smartphones has become an integral part of our daily lives. In fact, the number of smartphone mobile network subscriptions worldwide reached almost 6.4 billion in 2022, and is forecast to exceed 7.7 billion by 2028~\cite{number_of_smartphones}. Indeed, this large number of subscribers can be justified by the evolution of cellular technologies.
The latest, 5th Generation of Cellular Networks (5G) offers faster data speeds, lower latency, and increased capacity than its predecessors, enabling innovations such as the Internet of Things (IoT), massive MIMO and Millimeter Wave and Terahertz Communications~\cite{5G_ROADMAP,5G_paper_survey}.
%
However, the fact that smartphones and cellular networks have become a fundamental aspect of our lives raises an important question: \textit{How do we protect the privacy of cellular network users?}
Vast literature (see~\cite{cattaneo2013review,rupprecht2018security,vachhani2019security,cao2013survey} for some surveys) demonstrated how 4G/LTE networks and their predecessors (2G, 3G) are vulnerable to various privacy attacks for the User Equipment (UE).
Specifically, attacks were shown to violate different properties of UE Identity Confidentiality (Sec.~5.1 of~\cite{3g_security,3gpp_IMSI_catching_problem}), such as identity disclosure~\cite{paget2010practical,dabrowski2014imsi,kotuliak2022ltrack},  and location privacy~\cite{kim2019touching,hong2018guti,bitsikas2023freaky}.
Some attacks were demonstrated in 5G early versions~\cite{palama2021imsi,hussain20195greasoner} and even with cheap commodity equipment and open-source code implementations, thus, increasing the privacy risks since access to such tools can be easy. 

Indeed, it is a challenge for the current and future 5G networks to improve or mitigate the pre-5G networks' vulnerabilities, therefore, enhancing user privacy~\cite{liu2018toward,kumar2018user,rupprecht2018security,ahmad2018overview,jover2019current,khan2019survey,khan2020survey}.
Consequently, the 3rd Generation Partnership Project (3GPP), the international body that is responsible for cellular networks standardization, has considered the aforementioned problems after 5G Release 15, thus, introducing significant changes in the related 5G Security Technical Specification~\cite{3gpp_fake_base_stations}.
That said, recent surveys on the topic~\cite{rupprecht2018security,ahmad2018overview,khan2019survey,ferrag2018security} include only theoretical ideas about 5G privacy mechanisms, since they were written before or concurrently with the older, 5G Release 15.
Other surveys focused on specific use cases, such as the application of Machine Learning algorithms in 5G Physical layer~\cite{tanveer2021machine}, the security of the Early Data Transmission 5G mechanism~\cite{segura20225g}, 
or the security of alternative computing paradigms (e.g.,~catalytic computing) in 5G~\cite{choudhary2019survey}. Finally,~\cite{khan2020survey,yu2021improving}  analyzed previous cellular generations' attacks and studied corresponding mitigation mechanisms offered by 5G.
In general, past papers on the topic were written in earlier 5G specifications, thus, not being informed about newer Releases of 3GPP documents, recent 5G measurement studies~\cite{lasierra2023european,nie2022measuring,yu2023secchecker} and new privacy attacks~\cite{kotuliak2022ltrack,gao2023your,karakoc2023never}. 

In fact, an in-depth study of current literature on this topic raised more unanswered questions:
1) What vulnerabilities did the attacks in past cellular generations exploit, and what were the privacy implications of each?
2) What mitigation mechanisms have been introduced in the new release of 5G to overcome such weaknesses?
3) What gaps still remain for future exploration in the topic of 5G Security?
4) How recent 5G specific attacks take advantage of them?
In order to answer these questions, we perform this systemization of knowledge (SoK) paper.
We first propose a framework to study the efficiency of current cellular generation security aspects, given well-studied adversaries demonstrated in past cellular generations, under specific user privacy objectives.
Second, we examine the resilience of new 5G security features up to Release 19, called 5G-Advanced, 
with respect to attacks on previous generations.
Third, we examine their adaptability in terms of operators' implementation, classifying them as `optional' or `mandatory'.
Fourth, we raise open questions and gaps for future work.

With this SoK, our contributions are as follows:
\vspace{-3mm}
\begin{itemize}
\setlength\itemsep{-2pt}
    \item We define a framework for analyzing privacy-related attacks, setting the User Identity Confidentiality defined in 3GPP as our privacy objective (Sec.~\ref{Sec:Methodology}).
    \item We describe $12$ existing pre-5G attacks (2G, 3G, 4G/LTE) that violate User Identity Confidentiality, and highlight $12$ vulnerabilities or weaknesses that lead to them (Sec.~\ref{Sec:PREVIOUS_GENERATIONS}).
    \item We discuss security enhancements of 5G and mention $10$ mitigation mechanisms that are proposed against the aforementioned attacks (Sec.~\ref{Sec:5G_SECURITY}). 
    \item We describe $7$ recent 5G attacks along with their corresponding $7$ vulnerabilities, and the degree they are mitigated by the existing 5G MMs (Sec.~\ref{Sec:New_5G_attacks}).
    \item  We give a didactic summary of our results from Sec.~\ref{Sec:PREVIOUS_GENERATIONS},~\ref{Sec:5G_SECURITY} and~\ref{Sec:New_5G_attacks} in Table.~\ref{Table:Attacks_Overview}.
    \item We discuss key takeaways of our work and outline future research directions (Sec.~\ref{Sec: Discussion}).
\end{itemize}


\section{Background}\label{Sec:Background}
In this section, first, we provide basic information about the Cellular Network infrastructure and its key entities and functionalities (Sec.~\ref{Sec:Infrastructure}).
Second, we discuss the properties of different UE identifiers (Sec.~\ref{Sec:Identifiers}) and UE Capabilities (Sec.~\ref{Sec:Theory_Device_Capabilities}).
Then, we cover the 5G registration procedure with its corresponding protocols and  mechanisms (Sec.~\ref{Sec:Registration_Section}). 
We also provide a brief analysis of the paging procedure (Sec.~\ref{sec:Theory_Paging}).

\subsection{5G Cellular Network Infrastructure}\label{Sec:Infrastructure}

Here, we give an overview of the Cellular Network infrastructure, referring to the main entities and their properties.

\noindent
\textbf{User Equipment (UE):} The UE consists of the Mobile Equipment (ME) and USIM card. It is used by consumers to access mobile services and applications. 

\noindent
\textbf{Radio Access Network (RAN):} The Radio Access Network  consists of the different Base Stations that are separated into different Tracking Areas (TAs). It is responsible for managing radio resources, enabling handovers, providing wireless connectivity, and facilitating communication between UEs and the Core Network. The RAN comprises base stations (e.g.,~gnBs in 5G), antennas, and various hardware and software components that enable radio communication. The establishment, maintenance, and release of radio connections between the UE and the RAN is facilitated by the RRC  (Radio Resource Control) protocol~\cite{3GPP_RRC}.

\noindent
\textbf{Core Network (CN):} The Core Network consists of different entities called Network Functions (NFs), that are responsible for a variety of network services and functionalities. For instance, mobility management, authentication, subscriber data management, session management, charging control and connection to external networks are, among others, some important services provided by the CN via these functions. The signaling procedures and messages between the UE and the CN, including functions like registration, authentication, and mobility management are facilitated by the Non-Access Stratum (NAS) protocol~\cite{3GPP_NAS}. 


\subsection{Identifiers in 5G Cellular Networks}\label{Sec:Identifiers}

There are various identifiers used in the Cellular Network, responsible for the identification of different entities participating in the system, and especially of the UE at hand.
Among the different generations of cellular networks, there are differences between their structure and their name, but their role has been similar. In general, they are divided into permanent and temporary identifiers. 

To begin with, the International Mobile Subscriber Identity (IMSI) and International Mobile Equipment Identity (IMEI) are the most important permanent identifiers.
The first is the identifier of the USIM card used by the UE.
This identifier helps the CN identify the specific UE through time and networks. It is used mainly for the initial authentication of the UE to the network and in the past generations 2G, 3G and 4G, it was submitted in plaintext over the wireless channel.
As for the IMEI, it is the identifier of the Mobile Equipment (ME), manufactured on the device during its production. Both of them are permanent, global and are considered as extremely sensitive in terms of privacy.
5G introduced Subscription Permanent Identifier (SUPI) in the clause 5.9.2 of the 
3GPP 5G technical specifications for security architecture and procedures~\cite{5GS_architecture}. While intuitively SUPI is the same as IMSI, as we will see in the next paragraph regarding temporary identifiers, there is an important difference in the initial authentication: SUPI must never be submitted plaintext, except for emergency cases, as denoted in the clause 5.2.5 of the same 3GPP 5G technical specifications~\cite{3gpp_fake_base_stations}. As for the IMEI, the terminology of the Permanent Equipment Identifier (PEI) is used in 5G, as analyzed in Sec.~6.4 of~\cite{Numbering}. 

The other important category of identifiers are the temporary ones. First, Subscription Concealed Identifier (SUCI) is introduced in 5G (clause 5.9.2a in~\cite{5GS_architecture}).
SUCI is an elliptic cryptography-based concealed version of SUPI that is constructed by the USIM card. Furthermore, the CN assigns a temporary identifier to the UE for the communication between the UE and the CN (e.g.,~Service Request, paging procedures).
In the past, many terminologies have been used for this:
Temporary Mobile Subscriber Identity (TMSI) in 2G and 3G,
Globally Unique Temporary Identity (GUTI) or TMSI in 4G and
5G Globally Unique Temporary Identity (5G-GUTI) in 5G.
For the rest of this paper, we denote it as TMSI for 3G and 4G, and 5G-GUTI for 5G.
5G-GUTI is defined in the clause 5.9.4 of~\cite{5GS_architecture}, where the reader can also find information about how 5G-GUTI is constructed.
Finally, the UE is also assigned a temporary identifier called Cell Radio Network Temporary Identity (C-RNTI) from the RAN, facilitating the communication between the UE and the RAN.

\subsection{UE Capabilities}\label{Sec:Theory_Device_Capabilities}

The capabilities of a UE can be separated into Core Network (CN) capabilities~\cite{3GPP_NAS} and Radio Access capabilities~\cite{3GPP_Radio_Access_Capab}.
The CN capabilities indicate general UE characteristics, such as the security algorithms supported by the UE for integrity and ciphering protection, and are transmitted as a NAS message.
The Radio Access capabilities contain information regarding the radio capabilities of the UE, such as the supported frequency bands of the UE, and are transmitted as an RRC message. As explained in the following section, the UE reports its capabilities to the network during the registration procedure.
\cite{shaik2015practical,shaik2019new} provides further information about UE capabilities.


\begin{figure}
      \centering
  \includegraphics[width=0.99\linewidth]{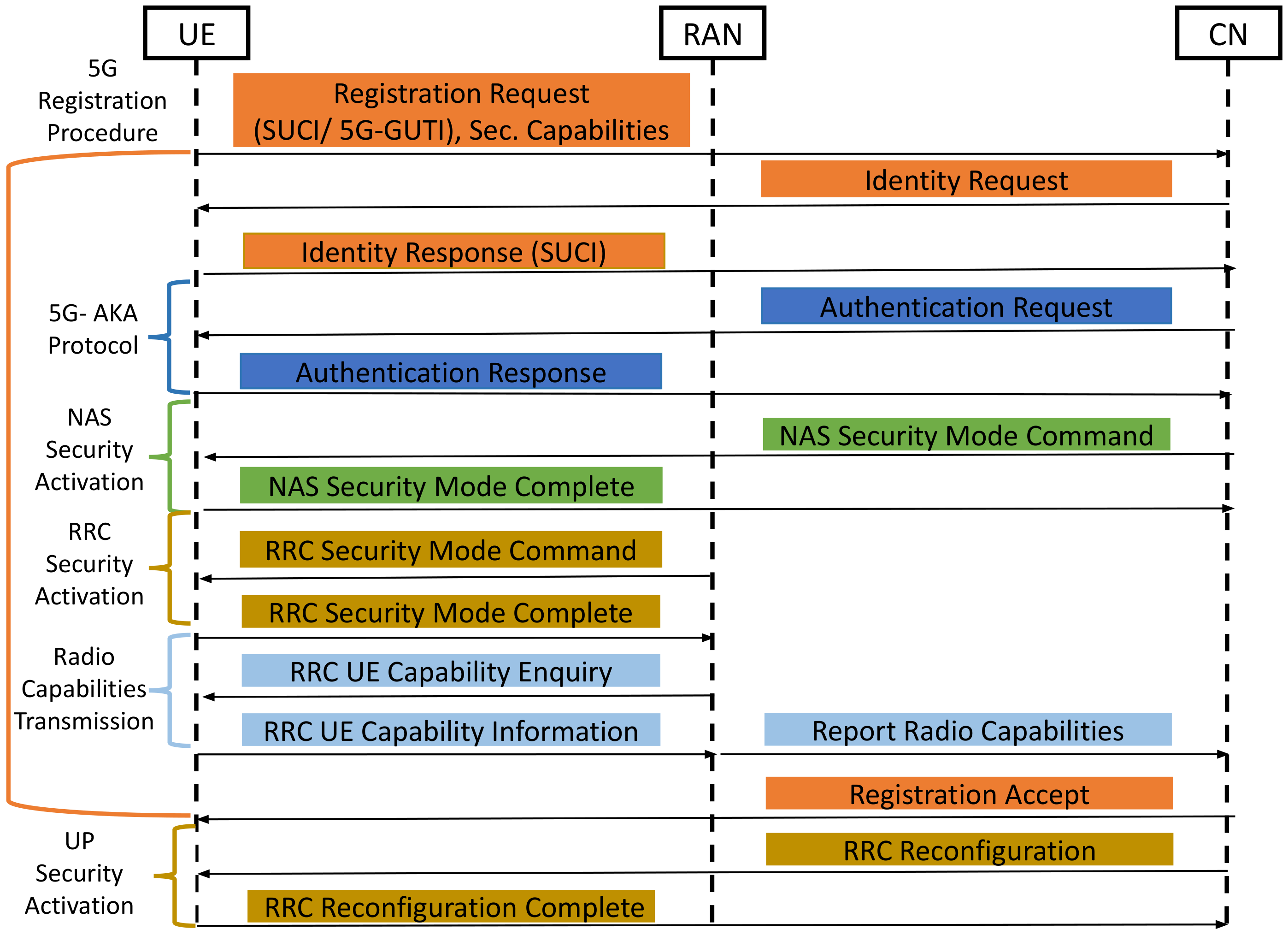} 
  \caption{UE Registration \& Authentication signal flow in 5G.}\label{fig:5G_flow}
  \vspace{-4mm}
\end{figure} 
\subsection{UE Registration \& Authentication Procedure}\label{Sec:Registration_Section}

Figure~\ref{fig:5G_flow} illustrates the basic message exchange flow with regards to some critical procedures in 5G, as introduced in the 3GPP 5G specifications~\cite{3gpp_fake_base_stations}. As shown in the figure, the UE sends a registration request including a subscriber identity (SUCI or 5G-GUTI) and the security capabilities. 
If the 5G-GUTI was sent and the CN cannot resolve it, it sends to the UE an Identity Request message. After that, the UE sends the SUCI in a Identity Response, and then the Authentication and Key Agreement Protocol (AKA) is initiated by the CN. We highlight that both EPS-AKA (Evolved Packet System Authentication and Key Agreement) and 5G-AKA (5G Authentication and Key Agreement) can be used, but since the differences are out of scope for this paper, 5G-AKA (Sec.~6.1.3.2 of~\cite{3gpp_fake_base_stations}) is used in this work.
In a nutsell, the network sends a random number (RAND) and an authentication token (AUTN) to the UE. The UE first verifies the validity of RAND and the freshness of AUTN and if the verification is not successful, it sends a MAC failure or Sync Failure respectively. If both of them are verified successfully, the UE uses the RAND and its permanent key to generate a response (RES) and sends it to the network as its Authentication Response.
The network verifies the validity of RES and, if it is valid, the authentication is successful. Further information about the 5G-AKA protocol can be found in~\cite{wang2021privacy,basin2018formal,cremers2019component,edrisformal_5GAKA}. After a successful authentication response sent by the UE, NAS Security Command is transmitted in order to activate a secure channel for the NAS protocol messages, providing integrity and ciphering protection. 

The same procedure is followed for the RRC messages, exchanged between the UE and the gnB, for the activation of a secure channel for them as well. Afterwards, the UE radio capabilities are transmitted to the 5G network, \textit{after the establishment of a secure channel} (see Figure~\ref{fig:5G_flow}). This is in contrast to prior cellular generations, where radio capabilities were sent before the establishment of a secure channel between the EU and the radio access network. Finally, User Plane (UP) ciphering and integrity is activated through the RRC Reconfiguration message. 

\subsection{Paging Mechanism}\label{sec:Theory_Paging}

The purpose of the paging mechanism is to inform the recipient UE for incoming data transmissions or a phone call. Paging messages are mainly sent by the Base Station (RRC paging), by broadcasting the identity of the UE, that is the recipient of incoming data or a call on the paging channel. In case of incoming data or an SMS, the paging procedure is initiated and all UEs within the cell listen to the paging channel and react to a message if their identity is received. In case of a phone call, the same procedure takes place in a TA level. As explained later, the paging mechanism was a major privacy problem for many cellular generations, since IMSI was used for it, facilitating adversaries to steal the UE's permanent identity.

\begin{figure}
\begin{subfigure}{.23\textwidth}
  \centering
  \includegraphics[width=.9\linewidth]{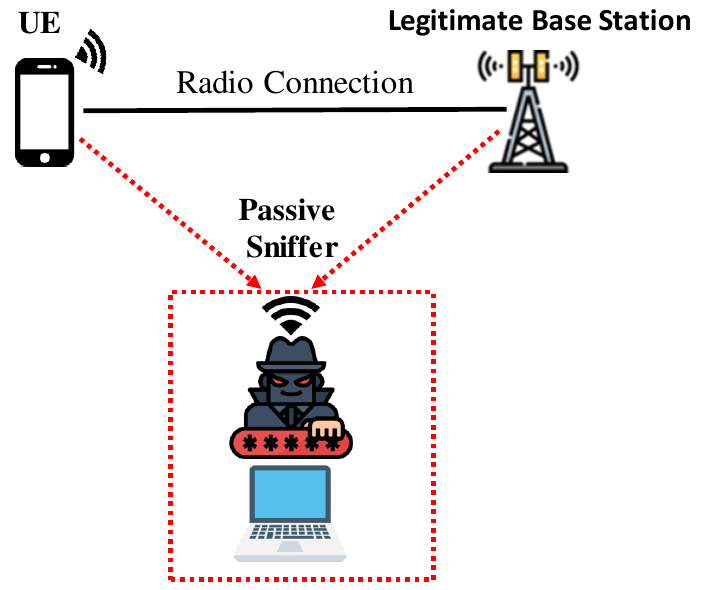}  
  \caption{Passive adversary.}
  \label{fig:passive_attack}
\end{subfigure}
\hfill
\begin{subfigure}{.23\textwidth}
  \centering
  \includegraphics[width=.9\linewidth]{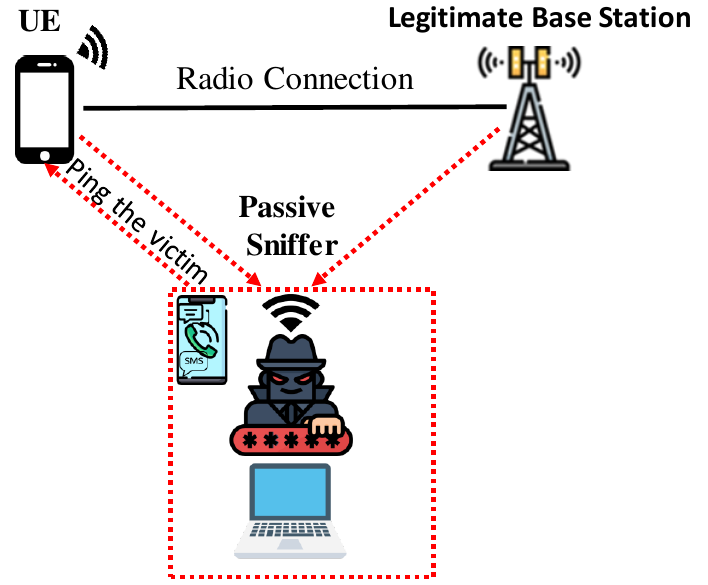}  
  \caption{Semi-passive adversary.}
  \label{fig:Semi_passive_attack}
\end{subfigure}
\hfill
\begin{subfigure}{.46\textwidth}
  \centering
  \includegraphics[width=.8\linewidth]{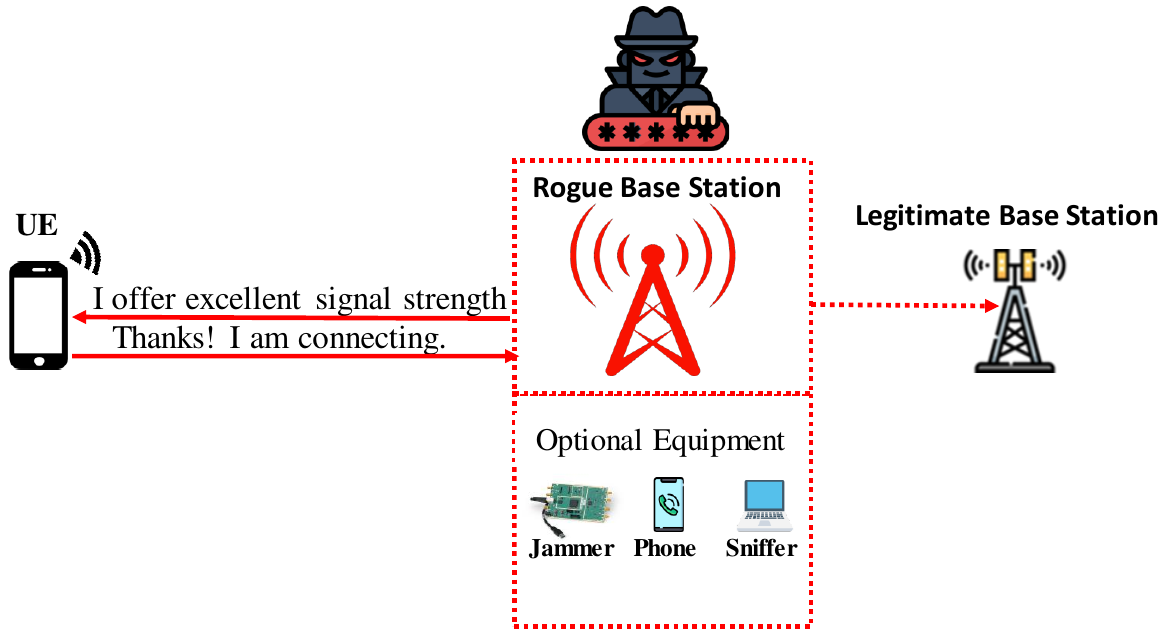}  
  \caption{Active adversary.}
  \label{fig:active_attack}
\end{subfigure}
\caption{Different types of adversaries in wireless networks.}
\label{fig:adversaries}
\vspace{-6mm}
\end{figure} 
\section{Methodology}\label{Sec:Methodology}

In this section, we describe the methodology of our work. 
First, in Sec.~\ref{Sec:privacy_objectives}, we define what User Identity Confidentiality is, and the properties that define it. Then, we give an overview of the different types of adversaries that try to violate the properties of User Identity Confidentiality (Sec.~\ref{Sec:Adversaries_type}).
Finally, Sec.~\ref{Sec:Framework} provides an overview of the Framework used for the systematization of the knowledge acquired by the various papers and other documents analyzed pertaining to this topic.

\subsection{User Identity Confidentiality}\label{Sec:privacy_objectives}

Taking into consideration the number of interconnected devices in today networks, and the wide range of sensitive data that are transmitted, it is important to define some privacy objectives about the confidentiality of the user identity. Starting from 3G and 4G cellular networks, the Sec.~5.1.1 of both~\cite{3g_security} and~\cite{3gpp_IMSI_catching_problem} refers to the security objective of the user identity confidentiality and its corresponding properties. In a nutshell, the protection of users' identity (IMSI), location and delivered services are of paramount importance. User identity confidentiality requirements are still relevant in 5G as stated in~\cite{basin2018formal}. For example, the Sec.~5.2.5 of~\cite{3gpp_fake_base_stations} refers to the user identity confidentiality, by mentioning  the SUPI and PEI protection, 5G related identifiers that are equal to IMSI and IMEI in previous generations as mentioned before in the Sec.~\ref{Sec:Identifiers}. In fact, UE identity confidentiality, as defined in the Sec.~5.1.1 of both~\cite{3g_security,3gpp_IMSI_catching_problem}, has been taken into consideration in numerous previous works~\cite{khan2020survey,khan2018identity,borgaonkar2018new,shen2022identity,khan2019efficacy}. Based on the above, UE identity confidentiality is considered as the privacy objective in this work and the following properties are necessary for its protection:

\begin{itemize}
\item UE Identity Privacy: "the permanent user identity (IMSI) of a user to whom a services is delivered cannot be eavesdropped on the radio access link".
\item UE Location Privacy: "the presence or the arrival of a user in a certain area cannot be determined by eavesdropping on the radio access link".
\item UE Untraceability: "an intruder cannot deduce whether different services are delivered to the same user by eavesdropping on the radio access link".
\end{itemize}

\subsection{Adversaries}\label{Sec:Adversaries_type}

We focus on types of adversaries reported in literature and classified as passive, semi-passive and active, as also proposed in~\cite{shaik2015practical}.
Next, we provide an overview for each (Fig.~\ref{fig:adversaries} illustrates each attack):
\begin{enumerate}
    \item Passive adversary: It has the ability of eavesdropping radio signals within a specific range, using channel sniffers~\cite{kumar2014lte}. It can receive, decode and store these radio signals, thereby managing to extract valuable and sensitive information for the UE. As it is passive, it is difficult to be detected. 
    \item Semi-Passive adversary: This is a passive adversary that can also somehow ping its target. The most common example is a passive adversary, that also sends some (silent) messages, or makes some (silent) calls~\cite{hong2018guti} to its victim.
    \item Active adversary: This adversary is capable of sending radio signals and messages to its target (e.g.,~\cite{lilly2017imsi,palama2021imsi}) or modify messages that were sent from or to the victim~\cite{karakoc2023never}. Most of the times, it is consisted of a fake base station that offers excellent signal strength~\cite{strobel2007imsi}. An active adversary may also have access to sniffers~\cite{bui2016owl}, another UE, or jammer devices~\cite{lichtman2016lte,lichtman20185g}. It is usually referred to as a ``Man-in-the-Middle'' (MiTM) adversary~\cite{yu2021improving}.
\end{enumerate}

We mention that nowadays, the equipment needed for such attacks can be both affordable and easily deployable. For example, a rogue base station can be constructed by using a Universal Software Radio Peripheral (USRP)~\cite{USRP} with a modified code of open-source projects like OpenLTE~\cite{OpenLTE}, srsRAN~\cite{srsRAN,gomez2016srslte}, gr-LTE~\cite{gr-LTE,demel2015lte} and OAI~\cite{OAI,nikaein2014openairinterface}, or by using a shorter range base station called femtocell~\cite{FEMTOCELLS}. Furthermore, there are available open-source tools for channel sniffers, such as~\cite{ludant20235g,hoang2023ltesniffer,kumar2014lte,bui2016owl,falkenberg2019falcon}, that are capable of decoding downlink traffic. Finally, jammers can be implemented using commodity equipment~\cite{jammerlte,pirayesh2022jamming}.

\begin{figure}[t]
  \centering
  \includegraphics[width=.99\linewidth]
  {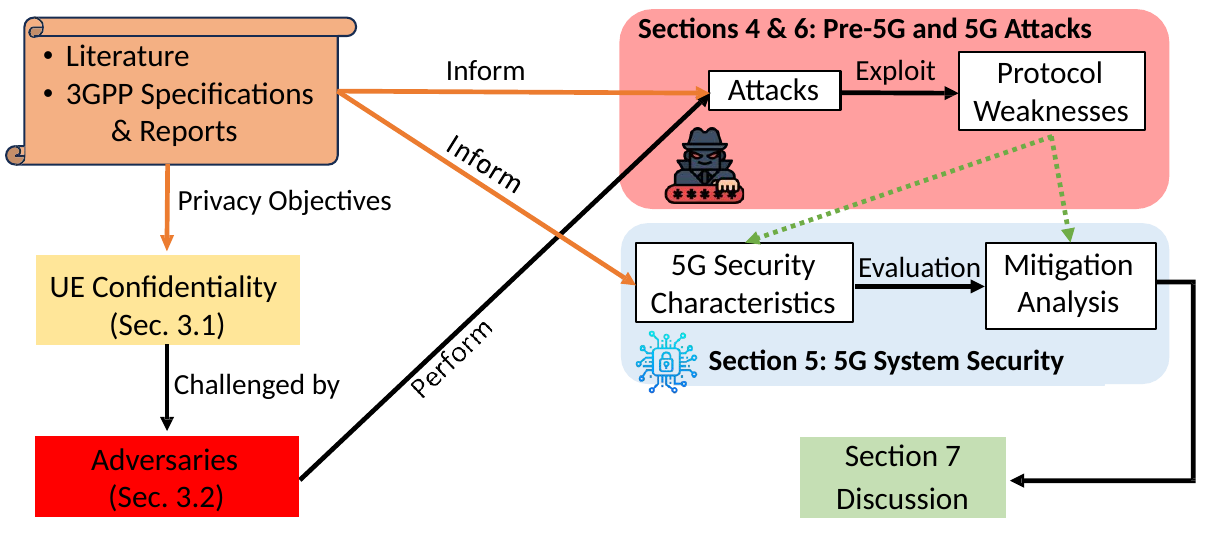} 
  \vspace{-2mm}
  \caption{General Framework of our Systematization.}
  \label{fig:framework}
\vspace{-6mm}
\end{figure}

\subsection{Framework}\label{Sec:Framework}

In this section, we explain the framework that is used for the systematization of the knowledge analyzed in this paper, as depicted in Figure~\ref{fig:framework}.
The very first step is to select the privacy requirements or objectives that should be met by the network system, in order to protect the different stakeholders. We study relevant scientific literature and 3GPP documents, in order to set as privacy objective the UE Identity Confidentiality, as analyzed earlier in Sec.~\ref{Sec:privacy_objectives}.
Then, we assume that the adversaries denoted in Sec.~\ref{Sec:Adversaries_type} try to challenge the UE Identity Confidentiality, by performing relevant attacks.
Thus, in the next Section~\ref{Sec:PREVIOUS_GENERATIONS}, we overview existing attacks mounted on 2G, 3G and LTE networks, obtaining information from the available literature.
During this process of attack review, we identify the weaknesses that lead to each attack, thus increasing our understanding for the contributing factors of each attack.
In the next step, as explained in Section~\ref{Sec:5G_SECURITY}, we refer again to the available scientific literature and 3GPP documents, but now focusing on 5G Systems (5GS) and their security characteristics. We first analyze these characteristics, compare them to the ones previously mentioned for the past cellular generations (Sec.~\ref{Sec:PREVIOUS_GENERATIONS}) and their corresponding privacy related weaknesses, and evaluate if these are mitigated in 5G; if yes, we also discuss to which degree. As mentioned before, the 2G, 3G, and 4G related attacks exploited the weaknesses we highlight.
Thus, it is straightforward that the (complete or partial) mitigation of these weaknesses contributes to the (complete or partial) mitigation of the related attacks as well in 5G. The same process is applied for recent 5G attacks in Sec.~\ref{Sec:New_5G_attacks}, highlighting their weaknesses and verifying if they can be mitigated by 5G mechanisms.

To facilitate the reader's understanding of the perspective of this work, we give a concrete example of our framework with the IMSI Catching.
This is a well-known attack (Sec.~\ref{Sec:IMSI_Cather}), performed by an active adversary, that violates all of the three properties of the UE Idenity Confidentiality, thus breaking the system's privacy objectives. The weakness behind this attack in 2G, 3G, and 4G is the \textit{plaintext transmission} of the USIM permanent identifier IMSI (UE identifiers were analyzed in Sec.~\ref{Sec:Identifiers}).
As a fix to this problem, a new 5G Security characteristic has been introduced, which is the SUCI mechanism (Sec.~\ref{SUCI_mechanism}), that encrypts the permanent USIM identifier. So, evaluating this new 5G security characteristic called SUCI, we conclude that the previously mentioned weakness, that led to IMSI catching attack, is mitigated.
Finally, in the last section, we refer to some limitations of this specific 5G security enhancement due to the lack of available 5G stand-alone (SA) networks, and not strict 3GPP regulations (SUCI is an optional feature).

\section{Pre-5G Cellular Generation Attacks}\label{Sec:PREVIOUS_GENERATIONS}

In this section, following our framework as defined in Sec.~\ref{Sec:Framework}, we analyze a variety of different attacks that are available in the scientific literature offering different layers of categorization. First, the different attacks are categorized into privacy and security attacks. Then, privacy attacks are further separated into permanent and temporary identifiers-based attacks, protocol exploitation attacks and measurement data exploitation attacks. Regarding  security attacks, they are categorized into Data Manipulation attacks and Protocol Downgrade ones. Further, we analyze the type of adversaries that can perform these attacks and evaluate their impact, in terms of UE Confidentiality as defined in Sec.~\ref{Sec:privacy_objectives}. Finally, the weaknesses that led to these attacks are outlined (in Italic font) and enumerated from $W1$ to $W12$. We summarize our findings on all attacks and weaknesses they take advantage of in Table~\ref{Table:Attacks_Overview}. 

\subsection{User Privacy attacks}

\subsubsection{\textbf{Attacks based on Permanent Identifiers}}\label{Sec:IMSI_attacks}
\paragraph{IMSI Catching:}\label{Sec:IMSI_Cather}
~IMSI Catching~\cite{meyer2004man,mitchell2001security,patent,lilly2017imsi,park2019anatomy,mjolsnes2019private,khan2008vulnerabilities}, which aims to steal the IMSI of the USIM, was one of the first practical attacks in 2G~\cite{perkov2011recent} and has also been persistent in 3G~\cite{ahmadian2009new} and 4G~\cite{michau2016not}. The attacker uses a device called IMSI Catcher, which as shown in~\cite{strobel2007imsi,van2016effectiveness,olimid2017lowcost,dubey2016demonstration} is easily deployable and affordable.
IMSI catchers operate in~\textit{active mode as fake (rogue) base stations}~\cite{dabrowski2014imsi,li2017fbs,palama2020diverse}, sometimes empowered by additional equipment, such as jammers~\cite{palama2021imsi}. First, the IMSI catcher takes advantage of the phone's behavior to connect to the Cell that offers the strongest signal power. Thus, the IMSI catcher is an active rogue base station that tries to make the victim UE connect to it, by offering better signal quality. When the UE connects to the IMSI catcher device, the adversary sends an Identity Request message. Then, the UE answers with an Identity Response message, including the IMSI without encryption (plaintext), thus, leading to UE identity disclosure, UE traceability and location tracking, consequences that break the UE Confidentiality, as defined in Sec~\ref{Sec:privacy_objectives}. Furthermore, IMSI catchers can work in conjunction with a variety of different attacks, such as SIM card cloning~\cite{liu2015small}, DoS attacks~\cite{mjolsnes2017easy,erni2022adaptover}, stealing of the UE's phone number~\cite{yu2019lte}.
Evidently, the plaintext IMSI transmission was characterized as a key vulnerability in the clause 6.1.3 of 3GPP Specifications~\cite{3gpp_IMSI_catching_problem}. When the IMSI of the UE has been obtained, the adversary can work in passive mode and only track the Person of Interest through their IMSI. The passive version described above is referred to as IMSI probing~\cite{2022imsiProbing} and has been characterized as low risk in 3GPP TS 33.846~\cite{imsiProbing3GPP}, so it is not analyzed further in this paper. 


Many different proposals have been made, aiming to stop IMSI Catchers. One one hand, there are five different sets of solutions working on the network side. A first set of solutions proposed either the usage of multiple IMSIs~\cite{khan2015improving,multipleIMSIs} per UE, or a new pseudonym instead of the IMSI~\cite{van2015defeating,norrman2016protecting,khan2018defeating}.
However, both of them suffer from synchronization problems between the USIM and the network as described in~\cite{khan2017trashing,hussain2018lteinspector}.
Second,~\cite{ekene2016enhanced,zhang2005security,deng2009novel,choudhury2012enhancing,geir2013privacy,li2011security} proposed significant changes both in the AKA protocol messages and the entities of the mobile network, thus making their implementations impractical.
In addition,~\cite{dabrowski2014imsi,dabrowski2016messenger,steig2016network,karaccay2021network,alrashede2019imsi} proposed network-based solutions, examining possible abnormalities (e.g.,~strange frequencies, unusual Cell locations and Cell IDs, signal noise level, unusual network parameters) that could have been created by the existence of an IMSI Catcher. It is unclear if any operator has implemented such a detection framework.
Besides,~\cite{van2015detecting,do2016strengthening,al2023cybersecurity} showed the feasibility of applying Machine Learning techniques for IMSI catcher detection but neither an exact framework was proposed nor issues like computational complexity and accuracy were analyzed. Last but not least, SeaGlass~\cite{ney2017seaglass} is a sensor-based approach where vehicles with portable sensors collect network measurements over a long period of time, trying to find anomalies caused by the presence of IMSI catchers. On the other hand, different applications were released~\cite{app1,app2,app3,app4,app5}, aiming to solve the problem from the subscribers' side, but they faced limitations such as  the low IMSI Catcher detection accuracy and the need of rooting permission to the user's phone, as analyzed in~\cite{park2017white,brenninkmeijer2016catching,ziayi2021yaicd}.
Summarizing, IMSI catching takes advantage of Weakness \#1 (W1):
 \setlength{\leftbarwidth}{4pt}
 \setlength{\leftbarsep}{1pt}
 \colorlet{leftbarcolor}{red}
\vspace{-1mm}
\begin{leftbar}
\noindent
\textbf{W1:}~\textit{The plaintext IMSI transmission during UE authentication.}
\end{leftbar}
\vspace{-1mm}

\paragraph{IMSI Extractor and UE Localization:}\label{Sec:IMSI_Extractor}
~The goal of the signal overshadowing attack is to disrupt the wireless communication by replacing legitimate signals over the air.
This attack requires time and frequency synchronization with the legitimate Base Station (BS), offering signal strength slightly stronger~\cite{yang2019hiding} or slightly weaker~\cite{ludant2021sigunder} than the legitimate one. In fact, this attack is stealthier compared to the traditional fake BS ones, since it uses a normal signal strength, thus making its detection even more difficult. Although most of the existing works use signal overshadowing for Denial of Service (DoS) attacks~\cite{yang2019hiding,ludant2021sigunder,erni2022adaptover}, a recent work~\cite{kotuliak2022ltrack} implements it for IMSI catching and UE passive localization. LTrack~\cite{kotuliak2022ltrack} is an adversary that uses a fake BS with slightly increased signal strength and perfect synchronization with the legitimate BS and just sends one adversarial message (Identity Request) to the victim. Then, the attacker uses an UL passive sniffer to record the Identity Response message, thus extracting the plaintext IMSI. This attack is called IMSI Extractor~\cite{kotuliak2022ltrack,erni2022adaptover}, in order to be differentiated by the traditional IMSI Catchers that do not use signal overshadowing techniques.  
Based on the definitions given in Sec.~\ref{Sec:Adversaries_type}, this adversary is active but if we consider that it transmits only one adversarial message, without neither increasing the signal strength of its rogue BS nor establishing a connection with the victim, there is a difference with the majority of existing attacks.
Since the target of this attack is accomplished mainly with the Uplink (UL) and Downlink (DL) sniffers, we characterize this attack as passive in order to show how stealthy is compared to the traditional rogue BS attacks. Finally, the adversary records the timing advance (TA), a parameter that is transmitted without encryption~\cite{roth2017location} in LTE, thereby managing to locate the victim with a localization error of around 6 meters. This is a standard weakness due to the message flow in 3G and LTE, since the TA is transmitted before the PCDP layer so the encryption has not been activated yet. On the contrary as stated in~\cite{rothlocation_case_study}, in 2G the TA eavesdropping problem did not exist since the signal flow was different and the TA was transmitted after the activation of encryption. For this reason, this attack is considered as partially applicable in 2G caused only by the plaintext IMSI transmission. Concluding, the two weaknesses behind this attack are W1 and a new, Weakness \#2 (W2):
\vspace{-1mm}
\begin{leftbar}
\noindent
\textbf{W2:} \textit{The lack of ciphering in MAC messages.}
\end{leftbar}
\vspace{-1mm}


\paragraph{IMSI Paging:}\label{Sec:IMSI_paging_problem}
~Another important problem comes from the paging procedures. 
Paging (Sec.~\ref{sec:Theory_Paging}) is initiated when the network searches for a UE in order to deliver a service to it, such as a phone call or an SMS. In general, the temporary identifier (TMSI) is used for paging, but in some cases (e.g.,~TMSI cannot be resolved by the network) IMSI can be used as well~\cite{raza2018exposing}. The fact that IMSI could be sent cleartext in paging messages made the paging process vulnerable to many semi-passive adversaries in 2G~\cite{kune2012location,golde2013let,kuklinski2020evaluation}, 3G~\cite{arapinis2012new,arapinis2017analysis} and 4G~\cite{shaik2015practical,bojic2017opportunities}. In this type of attack, the adversary already knows some information about the victim UE (e.g.,~phone number or social network accounts) and tries to take advantage of the paging process weaknesses to track the victim's location and also learn the UE's IMSI. The attacker initiates the paging process by sending (silent) messages, or making (silent) calls to the victim and at the same time they use a sniffer to observe the unencrypted downlink paging messages to identify the IMSI of the victim's UE. More information about silent calls and messages can be found in~\cite{hong2018guti,SILENT_voice_call}, but in summary, it is a call or message that activates the paging mechanism without the recipient to get notified. This attack violates the whole set of User Identity's Confidentiality properties, since it catches the IMSI (identity disclosure) and through the paging messages sniffing, also breaks the location and untraceability privacy, as well. Based on the above, the cause behind this attack is Weakness \#3 (W3):
\vspace{-1mm}
\begin{leftbar}
\noindent
\textbf{W3:}~\textit{Paging procedure with plaintext IMSI transmission.}
\end{leftbar}
\vspace{-1mm}


\paragraph{ToRPEDO:}\label{Sec:ToRPEDO}
Another problem of previous generations of cellular networks is that the Paging Occasions (POs) of a UE are correlated to the IMSI~\cite{hussain2019privacy,2020protecting_paging_PETS}. A PO is a specific time interval during which a UE is expected to monitor the paging channel for incoming paging messages. More specifically, one of the parameters of the PO is the Paging Frame Index (PFI) of a UE that is estimated using the IMSI ($PFI = IMSI \mod 1024$) (Sec.~7 of~\cite{3GPP_36304_LTE_paging}). The ToRPEDO semi-passive adversary~\cite{hussain2019privacy} observes the unencrypted and fixed PFI of a UE, thereby managing to obtain synchronization between the UE and its corresponding paging delay. Finally, ToRPEDO obtains information about the IMSI from the fixed PFI managing to learn 7 bits of the victim's IMSI, leading to partial User identity leakage among with location tracking and traceability. In a nutshell, the vulnerability that leads to this attack is Weakness \#4 (W4):
\vspace{-1mm}
\begin{leftbar}
\noindent
\textbf{W4:} \textit{Estimate of fixed POs based on IMSI}.
\end{leftbar}
\vspace{-1mm}



\paragraph{IMEI Catching:}\label{IMEI_Request} 
As mentioned earlier in Sec.~\ref{Sec:Identifiers}, IMEI is another sensitive permanent identifier, corresponding to the Mobile Equipment (ME). In 2G and 3G, the cleartext transmission of this identifier was permitted as a response to an Identity Request Message. Thus, an active adversary using a fake base station, similar to IMSI catchers' adversaries, could send an Identity Request using the IMEI instead of the IMSI, and steal the IMEI of the ME~\cite{dabrowski2016messenger,dabrowski2014imsi}. 
LTE standard specification identified this problem and changed the standard appropriately so that \textit{IMEI can be sent only after the activation of a secure channel}, as mentioned in~\cite{3gpp_IMSI_catching_problem} and~\cite{olimid2017lowcost}.
The analysis made in~\cite{mjolsnes2017experimental} verified the privacy enhancement of LTE in terms of IMEI privacy and as shown in~\cite{michau2016not,park2022doltest} only some old 4G MEs are vulnerable to the IMEI Catching attack. Based on the fact that this vulnerability in 4G/LTE comes from mistaken implementation in the ME side and not protocol vulnerability, IMEI catching is considered as solved in LTE. Concluding, the cause of this attack is Weakness \#5 (W5):
\vspace{-1mm}
\begin{leftbar}
\noindent
\textbf{W5:} \textit{Plaintext IMEI transmission.}
\end{leftbar}
\vspace{-1mm}

\subsubsection{\textbf{Attacks based on Temporary Identifiers}}
\hfill \break

\paragraph{TMSI Anonymity:}\label{Sec:guti_problems}
~The main reason for using the temporary identifier (TMSI) is the minimization of IMSI transmissions, offering better anonymity to the UE. In theory, TMSI has to be periodically updated by the network to avoid UE be easily tracked and identified~\cite{arapinis2014privacy}.
However, as shown in~\cite{arapinis2012new,arapinis2014privacy} the TMSI remained constant even for three days, in 2G and 3G networks during experiments that took place in different European countries. Similar results were obtained in LTE networks~\cite{shaik2015practical,hong2018guti,sorseth2019experimental}. More in detail, in~\cite{hong2018guti} the evidence for the mistaken GUTI refreshment rules is really strong. Detailed experiments in $11$ countries and $28$ different operators showed that even when the TMSI value was refreshed, the new value was predictable. A semi-passive adversary, consisting of a passive sniffer and a phone that sends some (silent) calls or messages to the victim's UE, leads to linkability of the victim's phone number with its TMSI. As a consequence, location tracking and traceability in 2G~\cite{kune2012location,saharan2017exploiting}, 3G~\cite{hong2018guti} and LTE~\cite{hong2018guti,shaik2015practical} was achieved, thus breaking two of the three privacy objectives defined in Sec.~\ref{Sec:privacy_objectives}. The main weakness behind TMSI Deanonymity attack is Weakness \#6 (W6):
\begin{leftbar}
\noindent
\textbf{W6:} \textit{Not frequent or missconfigured update policy of TMSI}.
\end{leftbar}
\vspace{-1mm}

\noindent


\paragraph{C-RNTI tracking:~}\label{Sec:RNTI_attacks}
C-RNTI based attacks is another potential danger for the UE location privacy~\cite{jover2016lte,jover2016lte_first,rupprecht2019breaking,kohls2019lost,oh2024enabling,bang2021impact,yu2012non} and untraceability~\cite{rupprecht2020call}. C-RNTI is local to the users’ serving Base Station (BS) and is used for the communication between the UE and the RAN. C-RNTIs can be found in both UL and DL control plane messages.~\cite{jover2016lte} passively analyzed the traffic in LTE and found that the C-RNTI is included without encryption in the header of every single packet, regardless of whether it is signaling or user traffic. 
Linkability between C-RNTI and the victim's phone number or a social network account (e.g.,~Telegram or WhatsApp) can be easily done by a semi-passive adversary with some silent messages or calls, as described in~\cite{hong2018guti}. In terms of decoding the DL messages including the C-RNTI, passive sniffers are available~\cite{falkenberg2019falcon,bui2016owl,kumar2014lte}, thus, breaking the UE location privacy and untraceability properties. The main vulnerability behind this attack is Weakness \#7 (W7):
\vspace{-1mm}
\begin{leftbar}
\noindent
\textbf{W7:} \textit{The lack of ciphering in RRC messages}.
\end{leftbar}
\vspace{1mm}

\subsubsection{\textbf{Protocol Exploitation attacks}}
\hfill \break

\paragraph{AKA Protocol Linkability:}\label{Sec:AKA_Linkability_Attack}
As mentioned in Sec.~\ref{Sec:Registration_Section}, Authentication and Key Agreement (AKA) protocols are used for the mutual authentication between the UE and CN from 3G and beyond. When the authentication is not successful, there are two failure reasons: MAC Failure or Sync Failure.
An active adversary~\cite{arapinis2012new,arapinis2017analysis,khan2014another} first observes an AKA session of the target user and records the Authentication Request including the authentication challenge (RAND) and token (AUTN). Then, the RAND and AUTN can be replayed by the adversary each time it wants to verify the presence of the target in a specific area. Indeed, thanks to the unencrypted failure messages (Sync or MAC failure), the adversary can distinguish other users from the target user who is the one the Authentication Request was originally sent to. The answer of the target user on the replayed (RAND, AUTN) is a Sync failure, whereas all the other users answer with MAC failure. Thus, linking two AKA sessions compromises the UE location privacy and traceability. We highlight that both 3G-AKA and 4G-AKA (EPS-AKA) did not solve the problem of linkability of AKA failure messages, since failure messages continue being transmitted in  cleartext~\cite{hahn2014privacy,karim2021prochecker}. Solutions to the AKA linkability problems were proposed both in 3G~\cite{arapinis2012new,arapinis2017analysis} and 4G~\cite{li2011security,hahn2014privacy,fei2023vulnerability}, but their compatibility with current deployments is limited due to wide changes to the AKA protocols, high computational overhead and high communication cost.
In a nutshell, this attack exploits Weakness \#8 (W8):
\vspace{-1mm}
\begin{leftbar}
\noindent
\textbf{W8:} \textit{AKA session failure cause is exposed in plaintext}.
\end{leftbar}
\vspace{-1mm}

\paragraph{Device Fingerprinting:}\label{Device_fingerprinting}
Device fingerprinting attack was proposed in LTE networks by Shaik et.~all~\cite{shaik2019new}. The adversary constructed a database of different devices' Core Network and Radio capabilities. They propose both a passive and an active version for their attack. First, a passive adversary eavesdrops on the first NAS message which up to LTE included the whole set of Core Network Capabilities (e.g., security algorithms supported, telephony features, power saving features, etc). Then the adversary uses the fingerprinting database to understand the device model and the service delivered to the user. Besides, an active version of this adversary transmits a UE Capabilities Inquiry message to the victim, exploiting the Weakness 12 as described later in the Sec.~\ref{Sec:Radio_Capab_attack}, and thus cheating the Radio Capabilities of the UE. As a result, the device fingerprinting attack is more accurate at this time since the adversary obtained both the Core Network and the Radio Capabilities of the UE. This attack breaks the location and untraceability properties and potentially the identity disclosure property as well, since the IMSI is transmitted in the first NAS message. The causes behind this attack is the W12 and a new one, Weakness \#9 (W9):
\vspace{-1mm}
\begin{leftbar}
\noindent
\textbf{W9:}~\textit{Transmission of the whole set of Core Network capabilities in the initial NAS message.}
\end{leftbar}
\vspace{2mm}

\subsubsection{\textbf{Measurement Data Exploitation Attacks\newline}}

\paragraph{UE measurement reports tracking:}\label{Sec:UE_Measurements_attack}
In cellular networks, UE performs network measurements and sends them to the Base Station (BS) when requested as an RRC message~\cite{saeed2022comprehensive,svensson20155g}.
Specifically, UE measurements report includes signal characteristics (e.g.,~signal strength of nearby BSs) facilitating the handover procedure and the maintenance of Radio Access Networks. An active attacker using a rogue base station with high signal quality can make the victim to connect to it, and ask for the UE measurements, managing to estimate the UE's location with triangulation~\cite{forsberg2007enhancing,shaik2015practical}. Furthermore,~\cite{olimid2017lowcost} proposes a passive version of this attack, by simply decoding the plaintext RRC messages including the UE measurement reports among with the user's C-RNTI. The above-mentioned papers refer to 4G networks, but this kind of attack is similar to 2G and 3G~\cite{bitsikas2021don}. We mention that LTE standards mandated that the UE measurement reports should be transmitted after the activation of an RRC secure channel in the Release 13 (LTE advanced) of TS.~36.331~\cite{RRC_3GPP_ciphering} and as a result, the active version of this attack is mitigated. On the other hand, the passive version of this attack can be mitigated only if the ciphering of RRC messages is activated, something that is operator's specific since the ciphering of RRC messages is optional~\cite{3gpp_IMSI_catching_problem}. Based on the above the UE measurements reports attack is considered as partially applicable in LTE networks as shown in Table~\ref{Table:Attacks_Overview}. The vulnerabilities behind this problem are W7 and a new one, Weakness \#10 (W10):
\vspace{-1mm}
\begin{leftbar}
\noindent
\textbf{W10 (up to LTE Release 13)}: \textit{Transmission of UE measurements reports before the establishment of a secure channel.}
\end{leftbar}

\subsection{User Security Attacks}

\subsubsection{\textbf{Data Manipulation Attacks\newline}}

\paragraph{ALTER and IMP4GT attacks:\label{Sec:Data_Manipulation}} 
 Data manipulation attacks are performed by active adversaries that exploit the lack of integrity in user plane messages, thus managing to redirect the victim UE to a malicious server or impersonate him~\cite{rupprecht2019breaking,rupprecht2020imp4gt}. More in detail, the ALTER attack~\cite{rupprecht2019breaking} allows to redirect a victim to a malicious website by manipulating the destination address of IP packets. The adversary intercepts and modifies the Uplink DNS request message, so that this message reaches an adversarial DNS server instead of the intended one. Then, the adversary also modifies accordingly the Downlink DNS response of the server, so that the attack is not noticed by the victim UE. Furthermore, IMPersonation in 4G neTworks (IMP4GT) attack~\cite{rupprecht2020imp4gt} is a dangerous extension of the ALTER attack that is able to impersonate not only the UE but the cellular network as well.
The cause of Data Manipulation attack is Weakness \#11 (W11):
\vspace{-1mm}
\begin{leftbar}
\noindent
\textbf{W11:}~\textit{Lack of integrity in user plane messages.}
\end{leftbar}
\vspace{-1mm}


\subsubsection{\textbf{Protocol Downgrade attacks\newline}}

\paragraph{Radio Capabilities Bidding-Down attack:}\label{Sec:Radio_Capab_attack} 
Bidding-Down attacks~\cite{shaik2019new,shaik2015practical,karakoc2023never} aim to downgrade the UE to a lower cellular technology. The lower the cellular generation the weaker the privacy mechanisms (e.g.,~2G has no AKA protocol), therefore, this kind of attack can potentially facilitate all the attacks mentioned in Sec.~\ref{Sec:PREVIOUS_GENERATIONS}.
More in detail,~\cite{shaik2019new,shaik2015practical} analyzes an attack based on the UE radio capabilities. An active adversary (rogue-base station) intercepts the UE Capability Enquiry message, including the radio capabilities of the UE, that was transmitted up to 4G before the establishment of a secure channel, so no integrity protection had been activated.
The adversary modifies appropriately the radio capabilities' characteristics (e.g.,~modify the frequencies supported by the UE Modem), thus managing to downgrade the UE to a lower cellular generation. Location privacy is directly violated since the adversary intercepts the victim's message, whereas identity privacy and untraceability properties may be potentially violated as well. The cause of this attack is Weakness \#12 (W12):
\vspace{-1mm}
\begin{leftbar}
\noindent
\textbf{W12:}~\textit{Transmission of UE Radio Capabilities before the establishment of a secure RRC channel.}
\end{leftbar}
\vspace{-1mm}

\newcommand{\tikzcircle}[2][black,fill=black]{\tikz[baseline=-0.5ex]\draw[#1,radius=#2] (0,0) circle ;}%
\newcommand*\halfcirc[1][0.1ex]{%
  \begin{tikzpicture}
  \draw[fill] (0,0)-- (90:#1) arc (90:270:#1) -- cycle ;
  \draw (0,0) circle (#1);
  \end{tikzpicture}}

\begin{table*}[]
\begin{tabular}{|l|ccc|ccc|cccc|cc|}
\hline
\multicolumn{1}{|c|}{\multirow{2}{*}{Attack Name}}                                  & \multicolumn{3}{c|}{\begin{tabular}[c]{@{}c@{}}Privacy \\ Implications\end{tabular}}                      & \multicolumn{3}{c|}{\begin{tabular}[c]{@{}c@{}}Adversary\\ Type\end{tabular}}           & \multicolumn{4}{c|}{\begin{tabular}[c]{@{}c@{}}Pre 5G\\ Generation\end{tabular}}                                                           & \multicolumn{2}{c|}{\begin{tabular}[c]{@{}c@{}}5G Security\\ Characteristics\end{tabular}}                                                                                                    \\ \cline{2-13} 
\multicolumn{1}{|c|}{}                                                              & \multicolumn{1}{c|}{\rotatebox{90}{Identity Disclosure}} & \multicolumn{1}{c|}{\rotatebox{90}{Location Tracking}} & \rotatebox{90}{Traceability}          & \multicolumn{1}{c|}{\rotatebox{90}{Active}} & \multicolumn{1}{c|}{\rotatebox{90}{Semi-Passive}} & \rotatebox{90}{Passive}               & \multicolumn{1}{c|}{\rotatebox{90}{2G}} & \multicolumn{1}{c|}{\rotatebox{90}{3G}} & \multicolumn{1}{c|}{\rotatebox{90}{4G}} & \begin{tabular}[c]{@{}c@{}}Weakness\\ Exploited\end{tabular} & \multicolumn{1}{c|}{\begin{tabular}[c]{@{}c@{}}Mitigation\\ Mechanisms\end{tabular}}                            & \begin{tabular}[c]{@{}l@{}} 
\\\rotatebox{90}{Optional or Mandatory}\end{tabular}\\ \hline
\begin{tabular}[c]{@{}l@{}}IMSI Catching \\\cite{strobel2007imsi,mitchell2001security,patent,paget2010practical,meyer2004man,dabrowski2014imsi,palama2021imsi,olimid2017lowcost,ahmadian2009new,michau2016not}  \end{tabular}                                                                         & \multicolumn{1}{c|}{\tikzcircle{2pt}}                    & \multicolumn{1}{c|}{\tikzcircle{2pt}}                  &   \tikzcircle{2pt}                    & \multicolumn{1}{c|}{\tikzcircle{2pt}}       & \multicolumn{1}{c|}{\tikzcircle[black, fill=white]{2pt}}             &   \tikzcircle[black, fill=white]{2pt}                    & \multicolumn{1}{c|}{\tikzcircle{2pt}}   & \multicolumn{1}{c|}{\tikzcircle{2pt}}   & \multicolumn{1}{c|}{\tikzcircle{2pt}}   & W1 (Sec.~\ref{Sec:IMSI_Cather})                                                           & \multicolumn{1}{c|}{\begin{tabular}[c]{@{}c@{}}MM1 (Sec.~\ref{SUCI_mechanism})\\ MM10 (Sec.~\ref{FBS_detection})\end{tabular}}                                                                                        & \begin{tabular}[c]{@{}c@{}}O \\ O \end{tabular}                                                                           \\ \hline
\begin{tabular}[c]{@{}l@{}}IMSI Extractor and\\ Localization~\cite{kotuliak2022ltrack,erni2022adaptover}\end{tabular}           & \multicolumn{1}{c|}{\tikzcircle{2pt}}                    & \multicolumn{1}{c|}{\tikzcircle{2pt}}                  &  \tikzcircle{2pt}                     & \multicolumn{1}{c|}{\tikzcircle[black, fill=white]{2pt}}       & \multicolumn{1}{c|}{\tikzcircle[black, fill=white]{2pt}}             &           \tikzcircle{2pt}            & \multicolumn{1}{c|}{\halfcirc[0.4ex]}   & \multicolumn{1}{c|}{\tikzcircle{2pt}}   & \multicolumn{1}{c|}{\tikzcircle{2pt}}   & \begin{tabular}[c]{@{}c@{}}W1 (Sec.~\ref{Sec:IMSI_Cather})\\ W2 (Sec.~\ref{Sec:IMSI_Extractor})\end{tabular}              & \multicolumn{1}{c|}{\begin{tabular}[c]{@{}c@{}}MM1 (Sec.~\ref{SUCI_mechanism})\\ MM10 (Sec.~\ref{FBS_detection})\end{tabular}}                                 & \begin{tabular}[c]{@{}c@{}}O\\ O\end{tabular}                               \\ \hline
\begin{tabular}[c]{@{}l@{}}IMSI Paging 
~\cite{kune2012location,arapinis2017analysis,shaik2015practical,bojic2017opportunities}\end{tabular}                                                                           & \multicolumn{1}{c|}{\tikzcircle{2pt}}                    & \multicolumn{1}{c|}{\tikzcircle{2pt}}                  &  \tikzcircle{2pt}                     & \multicolumn{1}{c|}{\tikzcircle[black, fill=white]{2pt}}       & \multicolumn{1}{c|}{\tikzcircle{2pt}}             &  \tikzcircle[black, fill=white]{2pt}                     & \multicolumn{1}{c|}{\tikzcircle{2pt}}   & \multicolumn{1}{c|}{\tikzcircle{2pt}}   & \multicolumn{1}{c|}{\tikzcircle{2pt}}   & W3 (Sec.~\ref{Sec:IMSI_paging_problem})                                                           & \multicolumn{1}{c|}{MM3 (Sec.~\ref{5G_TMSI_solutions})}                                                                                           & M                                                                           \\ \hline
ToRPEDO~\cite{hussain2019privacy}                                                                             & \multicolumn{1}{c|}{\halfcirc[0.4ex]}                    & \multicolumn{1}{c|}{\tikzcircle{2pt}}                  &  \tikzcircle{2pt}                     & \multicolumn{1}{c|}{\tikzcircle[black, fill=white]{2pt}}       & \multicolumn{1}{c|}{\tikzcircle{2pt}}             & \tikzcircle[black, fill=white]{2pt}                      & \multicolumn{1}{c|}{\tikzcircle{2pt}}   & \multicolumn{1}{c|}{\tikzcircle{2pt}}   & \multicolumn{1}{c|}{\tikzcircle{2pt}}   & W4 (Sec.~\ref{Sec:ToRPEDO})                                                           & \multicolumn{1}{c|}{MM4 (Sec.~\ref{5G_TMSI_solutions})}                                                                                           & M                                                                           \\ \hline
\begin{tabular}[c]{@{}l@{}}IMEI Catching \\\cite{dabrowski2014imsi,olimid2017lowcost,michau2016not,park2022doltest}\end{tabular}        & \multicolumn{1}{c|}{\halfcirc[0.4ex]}                    & \multicolumn{1}{c|}{\tikzcircle{2pt}}                  & \halfcirc[0.4ex]                      & \multicolumn{1}{c|}{\tikzcircle{2pt}}       & \multicolumn{1}{c|}{\tikzcircle[black, fill=white]{2pt}}             & \tikzcircle[black, fill=white]{2pt}                      & \multicolumn{1}{c|}{\tikzcircle{2pt}}   & \multicolumn{1}{c|}{\tikzcircle{2pt}}   & \multicolumn{1}{c|}{\tikzcircle[black, fill=white]{2pt}}   & W5 (Sec.~\ref{IMEI_Request})                                                           & \multicolumn{1}{c|}{\begin{tabular}[c]{@{}c@{}}Solved since LTE\end{tabular}} & M                                                                           \\ \hline
\begin{tabular}[c]{@{}l@{}}TMSI Anonymity \\ \cite{kune2012location,arapinis2012new,arapinis2014privacy,hong2018guti}  \end{tabular}                                                                                                                                       & \multicolumn{1}{c|}{\tikzcircle[black, fill=white]{2pt}}                    & \multicolumn{1}{c|}{\tikzcircle{2pt}}                  &  \tikzcircle{2pt}                     & \multicolumn{1}{c|}{\tikzcircle[black, fill=white]{2pt}}       & \multicolumn{1}{c|}{\tikzcircle{2pt}}             & \tikzcircle[black, fill=white]{2pt}                      & \multicolumn{1}{c|}{\tikzcircle{2pt}}   & \multicolumn{1}{c|}{\tikzcircle{2pt}}   & \multicolumn{1}{c|}{\tikzcircle{2pt}}   & W6 (Sec.~\ref{Sec:guti_problems})                                                           & \multicolumn{1}{c|}{MM2 (Sec.~\ref{5G-GUTI_update})}                                                                                           & M                                                                           \\ \hline
\begin{tabular}[c]{@{}l@{}}C-RNTI tracking \\\cite{jover2016lte_first,jover2016lte,rupprecht2019breaking,rupprecht2020call} \end{tabular}                                                                        & \multicolumn{1}{c|}{\tikzcircle[black, fill=white]{2pt}}                    & \multicolumn{1}{c|}{\tikzcircle{2pt}}                  &  \tikzcircle{2pt}                     & \multicolumn{1}{c|}{\tikzcircle[black, fill=white]{2pt}}       & \multicolumn{1}{c|}{\tikzcircle{2pt}}             & \tikzcircle[black, fill=white]{2pt}                      & \multicolumn{1}{c|}{\tikzcircle{2pt}}   & \multicolumn{1}{c|}{\tikzcircle{2pt}}   & \multicolumn{1}{c|}{\halfcirc[0.4ex]}   & W7 (Sec.~\ref{Sec:RNTI_attacks})                                                          & \multicolumn{1}{c|}{MM6 (Sec.~\ref{Sec:Sol_integrity_confid})}                                                                                           & O                                                                           \\ \hline

\begin{tabular}[c]{@{}l@{}}AKA Protocol \\ Linkability~\cite{arapinis2012new,arapinis2017analysis,khan2014another,hahn2014privacy}\end{tabular}                 & \multicolumn{1}{c|}{\tikzcircle[black, fill=white]{2pt}}                    & \multicolumn{1}{c|}{\tikzcircle{2pt}}                  &  \tikzcircle{2pt}                     & \multicolumn{1}{c|}{\tikzcircle{2pt}}       & \multicolumn{1}{c|}{\tikzcircle[black, fill=white]{2pt}}             &  \tikzcircle[black, fill=white]{2pt}                     & \multicolumn{1}{c|}{\tikzcircle[black, fill=white]{2pt}}   & \multicolumn{1}{c|}{\tikzcircle{2pt}}   & \multicolumn{1}{c|}{\tikzcircle{2pt}}   & W8 (Sec.~\ref{Sec:AKA_Linkability_Attack})                                                           & \multicolumn{1}{c|}{No MM}                                                                                      & -                                                                           \\ \hline

Device Fingerprinting~\cite{shaik2019new}                                                               & \multicolumn{1}{c|}{\halfcirc[0.4ex]}                    & \multicolumn{1}{c|}{\tikzcircle{2pt}}                  &   \tikzcircle{2pt}                    & \multicolumn{1}{c|}{\tikzcircle{2pt}}       & \multicolumn{1}{c|}{\tikzcircle[black, fill=white]{2pt}}             &  \tikzcircle{2pt}                     & \multicolumn{1}{c|}{\tikzcircle{2pt}}   & \multicolumn{1}{c|}{\tikzcircle{2pt}}   & \multicolumn{1}{c|}{\tikzcircle{2pt}}   & \multicolumn{1}{c|}{\begin{tabular}[c]{@{}c@{}} W12 (Sec.~\ref{Sec:Radio_Capab_attack})\\ W9 (Sec.~\ref{Device_fingerprinting})\end{tabular}}                                                          & \multicolumn{1}{c|}{\begin{tabular}[c]{@{}c@{}}MM9 (Sec.~\ref{Secure_Radio_Capab})\\MM8 (Sec.~\ref{Solution_device_capab})\end{tabular}}                                                                                       & \multicolumn{1}{c|}{\begin{tabular}[c]{@{}c@{}} M \\ M \end{tabular}}                                                                            \\ \hline
\begin{tabular}[c]{@{}l@{}}UE Measurements\\ reports~\cite{forsberg2007enhancing,shaik2015practical,olimid2017lowcost,bitsikas2021don}\end{tabular}                   & \multicolumn{1}{c|}{\tikzcircle[black, fill=white]{2pt}}                    & \multicolumn{1}{c|}{\tikzcircle{2pt}}                  & \tikzcircle{2pt} & \multicolumn{1}{c|}{\tikzcircle{2pt}}       & \multicolumn{1}{c|}{\halfcirc[0.4ex]}             & \tikzcircle[black, fill=white]{2pt} & \multicolumn{1}{l|}{\tikzcircle{2pt}}   & \multicolumn{1}{l|}{\tikzcircle{2pt}}   & \multicolumn{1}{l|}{\halfcirc[0.4ex]}   &\begin{tabular}[c]{@{}l@{}} W7 (Sec.~\ref{Sec:RNTI_attacks}) \\
W10 (Sec.~\ref{Sec:UE_Measurements_attack})\end{tabular}                                                           & \multicolumn{1}{c|}{\begin{tabular}[c]{@{}c@{}}MM6 (Sec.~\ref{Sec:Sol_integrity_confid})\\ Solved since LTE \end{tabular}}                                                                                        &\begin{tabular}[c]{@{}c@{}} O \\ M\end{tabular}                                                                           \\ \hline
Data Manipulation~\cite{rupprecht2019breaking,rupprecht2020imp4gt}                                                                   & \multicolumn{1}{c|}{\tikzcircle[black, fill=white]{2pt}}                    & \multicolumn{1}{c|}{\tikzcircle{2pt}}                  & \tikzcircle{2pt}                       & \multicolumn{1}{c|}{\tikzcircle{2pt}}       & \multicolumn{1}{c|}{\tikzcircle[black, fill=white]{2pt}}             &  \tikzcircle[black, fill=white]{2pt}                     & \multicolumn{1}{c|}{\tikzcircle{2pt}}   & \multicolumn{1}{c|}{\tikzcircle{2pt}}   & \multicolumn{1}{c|}{\tikzcircle{2pt}}   & W11 (Sec.~\ref{Sec:Data_Manipulation})                                                           & \multicolumn{1}{c|}{\begin{tabular}[c]{@{}c@{}}MM7 (Sec.~\ref{Sec:Sol_integrity_confid})\\MM10 (Sec.~\ref{FBS_detection})\end{tabular}}                                                                                           &\begin{tabular}[c]{@{}c@{}}O \\ O \end{tabular}                                                                          \\ \hline
\begin{tabular}[c]{@{}l@{}}Radio Capabilities\\ bidding down \\ attack~~\cite{shaik2015practical,shaik2019new,karakoc2023never}\end{tabular} & \multicolumn{1}{c|}{\halfcirc[0.4ex]}                    & \multicolumn{1}{c|}{\tikzcircle{2pt}}                  &   \halfcirc[0.4ex]                    & \multicolumn{1}{c|}{\tikzcircle{2pt}}       & \multicolumn{1}{c|}{\tikzcircle[black, fill=white]{2pt}}             & \tikzcircle[black, fill=white]{2pt}                      & \multicolumn{1}{c|}{\tikzcircle[black, fill=white]{2pt}}   & \multicolumn{1}{c|}{\tikzcircle{2pt}}   & \multicolumn{1}{c|}{\tikzcircle{2pt}}   & W12 (Sec.~\ref{Sec:Radio_Capab_attack})                                                          & \multicolumn{1}{c|}{\begin{tabular}[c]{@{}c@{}}MM9 (Sec.~\ref{Secure_Radio_Capab}) \end{tabular}}                                                                                           &  \begin{tabular}[c]{@{}c@{}}M     \end{tabular}                                                                      \\ \hline
\hline
Quantum SUCI~\cite{SUCI_QUANTUM}                                                                        & \multicolumn{1}{c|}{\tikzcircle{2pt}}                    & \multicolumn{1}{c|}{\tikzcircle{2pt}}                  & \multicolumn{1}{c|}{\tikzcircle{2pt}} & \multicolumn{1}{c|}{\tikzcircle{2pt}}       & \multicolumn{1}{c|}{\tikzcircle[black, fill=white]{2pt}}             & \multicolumn{1}{c|}{\tikzcircle[black, fill=white]{2pt}} & \multicolumn{1}{c|}{}   & \multicolumn{1}{c|}{}   & \multicolumn{1}{c|}{}   & \multicolumn{1}{c|}{W13 (Sec.~\ref{Sec:suci_quantum})}                                        & \multicolumn{1}{c|}{MM5 (Sec.~\ref{Sec:Sol_integrity_confid})}                                                                                           & O                                                       \\ \hline
\begin{tabular}[c]{@{}l@{}}CSI Reports localization\\ attack~\cite{ludant2024unprotected} \end{tabular}                                                                 & \multicolumn{1}{c|}{\tikzcircle[black, fill=white]{2pt}}                    & \multicolumn{1}{c|}{\tikzcircle{2pt}}                  & \multicolumn{1}{c|}{\tikzcircle[black, fill=white]{2pt}} & \multicolumn{1}{c|}{\tikzcircle[black, fill=white]{2pt}}       & \multicolumn{1}{c|}{\tikzcircle{2pt}}             & \multicolumn{1}{c|}{\tikzcircle[black, fill=white]{2pt}} & \multicolumn{1}{c|}{}   & \multicolumn{1}{c|}{}   & \multicolumn{1}{c|}{}   & \multicolumn{1}{c|}{W14 (Sec.~\ref{CSI_reports_attack})}                                        & \multicolumn{1}{c|}{No MM}                                                                                       & -                                                       \\ \hline

  \begin{tabular}[c]{@{}l@{}}ISAC based tracking attack\\\cite{lu2024integrated,wang2022integrated,liu_sensing,liu2023exploring}\end{tabular}                   & \multicolumn{1}{c|}{\tikzcircle[black, fill=white]{2pt}}                    & \multicolumn{1}{c|}{\tikzcircle{2pt}}                  & \multicolumn{1}{c|}{\tikzcircle{2pt}} & \multicolumn{1}{c|}{\tikzcircle{2pt}}       & \multicolumn{1}{c|}{\tikzcircle[black, fill=white]{2pt}}             & \multicolumn{1}{c|}{\tikzcircle{2pt}} & \multicolumn{1}{c|}{}   & \multicolumn{1}{c|}{}   & \multicolumn{1}{c|}{}   & \multicolumn{1}{c|}{W15 (Sec.~\ref{Sec:ISAC_based_attacks})}                                        & \multicolumn{1}{c|}{No MM}                                                                                           & \multicolumn{1}{c|}{-}  
\\ \hline
\begin{tabular}[c]{@{}l@{}}Satellite and NTN tracking\\ attack~\cite{pavur2020sok,RECORD_ATTACK}\end{tabular}                   & \multicolumn{1}{c|}{\tikzcircle[black, fill=white]{2pt}}                    & \multicolumn{1}{c|}{\tikzcircle{2pt}}                  & \multicolumn{1}{c|}{\tikzcircle{2pt}} & \multicolumn{1}{c|}{\tikzcircle{2pt}}       & \multicolumn{1}{c|}{\tikzcircle[black, fill=white]{2pt}}             & \multicolumn{1}{c|}{\tikzcircle{2pt}} & \multicolumn{1}{c|}{}   & \multicolumn{1}{c|}{}   & \multicolumn{1}{c|}{}   & \multicolumn{1}{c|}{W16 (Sec.~\ref{Sec:NTN_based_attacks})}                                        & \multicolumn{1}{c|}{No MM}                                                                                           & \multicolumn{1}{c|}{-}  
\\ \hline
PRS spoofing attack~\cite{gao2023your,gao2024surgical}                                                                    & \multicolumn{1}{c|}{\tikzcircle[black, fill=white]{2pt}}                    & \multicolumn{1}{c|}{\tikzcircle{2pt}}                  & \multicolumn{1}{c|}{\tikzcircle[black, fill=white]{2pt}} & \multicolumn{1}{c|}{\tikzcircle{2pt}}       & \multicolumn{1}{c|}{\tikzcircle[black, fill=white]{2pt}}             & \multicolumn{1}{c|}{\tikzcircle[black, fill=white]{2pt}} & \multicolumn{1}{c|}{}   & \multicolumn{1}{c|}{}   & \multicolumn{1}{c|}{}   & \multicolumn{1}{c|}{W17 (Sec.~\ref{Sec:PRS_attack})}                                        & \multicolumn{1}{c|}{MM10 (Sec.~\ref{FBS_detection})}                                                                                           & O                                                       \\\hline

  \begin{tabular}[c]{@{}l@{}}Ambient-IoT device \\ spoofing attack~\cite{narayanan2021harvestprint,jiang2023backscatter}\end{tabular}                   & \multicolumn{1}{c|}{\tikzcircle[black, fill=white]{2pt}}                    & \multicolumn{1}{c|}{\tikzcircle{2pt}}                  & \multicolumn{1}{c|}{\tikzcircle{2pt}} & \multicolumn{1}{c|}{\tikzcircle{2pt}}       & \multicolumn{1}{c|}{\tikzcircle[black, fill=white]{2pt}}             & \multicolumn{1}{c|}{\tikzcircle[black, fill=white]{2pt}} & \multicolumn{1}{c|}{}   & \multicolumn{1}{c|}{}   & \multicolumn{1}{c|}{}   & \multicolumn{1}{c|}{W18 (Sec.~\ref{Sec:AmbientIoT_attack})}                                        & \multicolumn{1}{c|}{No MM}                                                                                           & \multicolumn{1}{c|}{-}  
\\ \hline
\begin{tabular}[c]{@{}l@{}}5G Bidding Down \\ attack~\cite{hussain20195greasoner}\end{tabular}                   & \multicolumn{1}{c|}{\halfcirc[0.4ex]}                    & \multicolumn{1}{c|}{\tikzcircle{2pt}}                  & \multicolumn{1}{c|}{\halfcirc[0.4ex]} & \multicolumn{1}{c|}{\tikzcircle{2pt}}       & \multicolumn{1}{c|}{\tikzcircle[black, fill=white]{2pt}}             & \multicolumn{1}{c|}{\tikzcircle[black, fill=white]{2pt}} & \multicolumn{1}{c|}{}   & \multicolumn{1}{c|}{}   & \multicolumn{1}{c|}{}   & \multicolumn{1}{c|}{W19 (sec.~\ref{Sec:5G_bidding_down_attack})}                                        & \multicolumn{1}{c|}{MM10 (Sec.~\ref{FBS_detection})}                                                                                           & \multicolumn{1}{c|}{O}                                                       \\ \hline
\end{tabular}

\vspace{1mm}{\tikzcircle{2pt}: Applicable,~\halfcirc[0.5ex]: Partially Applicable,~\tikzcircle[black, fill=white]{2pt}: Not Applicable.}
\caption{
Top part: Overview of pre-5G (2G-4G) attacks against UE Confidentiality, corresponding weakness (W\#) exploited, and any mitigation method (MM\#) proposed.
Bottom part: Overview of new 5G attacks, corresponding weakness exploited and any mitigation method proposed.
}
\label{Table:Attacks_Overview}
\end{table*}

\section{5G Systems Security}\label{Sec:5G_SECURITY}

In this section, we analyze the security protocols and mechanisms proposed for 5G in the related 3GPP document~\cite{3gpp_fake_base_stations}, aiming to minimize or eliminate the problems and security risks of Sec.~\ref{Sec:PREVIOUS_GENERATIONS}. These improvements stand as mitigation mechanisms, as mentioned in our Framework in Sec.~\ref{Sec:Framework} and are enumerated clearly in the end of each subsection. Further, we evaluate their potential efficiency in terms of privacy enhancement, matching them to the weaknesses ($W1-W12$) mentioned in Sec.~\ref{Sec:PREVIOUS_GENERATIONS}. Finally, their complete or partial adaptability (optional or mandatory mechanisms) to the current 5G networks is considered. The last two columns of Table~\ref{Table:Attacks_Overview} summarize our findings for 5G.

\subsection{Improved UE Identity Privacy based on SUCI}\label{SUCI_mechanism} 

As mentioned in Sec.~\ref{Sec:Identifiers}, SUPI is the corresponding Identifier to IMSI in 5G networks. In order to avoid the plaintext transmission of SUPI, Subscriber Unique Concealed Identifier (SUCI) is introduced as an~\textit{encrypted} form of SUPI, based on elliptic cryptography. In fact, SUCI can be mainly used for authentication if the temporary identifier, 5G-GUTI, is not available. We highlight that the implementation of SUCI is of paramount importance for the UE's privacy, since it mitigates the attacks imposed by the IMSI plaintext transmission as described in Sections~\ref{Sec:IMSI_Cather} and~\ref{Sec:IMSI_Extractor}.
Therefore, we understand that SUCI mechanism can really improve UE identity privacy. However, its implementation is still optional based on~\cite{3gpp_fake_base_stations}.
In summary, the mitigation mechanism \#1 (MM1) is:
\setlength{\leftbarwidth}{4pt}
\setlength{\leftbarsep}{1pt}
\colorlet{leftbarcolor}{blue}
\vspace{-1mm}
\begin{leftbar}
\noindent
\textbf{MM1:}~\textit{Concealment of SUPI using SUCI.}
\end{leftbar}
\vspace{-1mm}


\subsection{Improvements on 5G-GUTI} \label{5G-GUTI improvements}

\subsubsection{Strict 5G-GUTI update mechanism} \label{5G-GUTI_update}

5G Global Unique Temporary Identifier (5G-GUTI) is the corresponding identifier to TMSI in previous cellular network generations. As analyzed in Sec.~\ref{Sec:guti_problems}, previous generations faced serious privacy problems due to the infrequent or miss-configured refreshment of this temporary identifier~\cite{hong2018guti}. Based on this outlook, 5G strictly defines when the 5G-GUTI should be updated or refreshed by the Core Network in the clause 6.12.3 of~\cite{3gpp_fake_base_stations}:
\begin{itemize}
\setlength{\itemsep}{-2pt}
    \item Upon receiving Registration Request message of type "initial registration" or "mobility registration update" from UE.
    \item Upon receiving Service Request message sent by the UE in response to a Paging message.
    \item Upon receiving Registration Request message of type "periodic registration update" from a UE.
    \item Upon receiving an indication from lower layers the RRC connection has been resumed for a UE in 5GMM IDLE mode with suspend indication in response to a Paging message.
    \item Even more frequently, based on operator implementation.
\end{itemize}
In addition, it is denoted in the same 3GPP document that 5G-GUTI should be generated in an \textit{unpredictable} way, bringing us to the Mitigation Mechanism \#2 (MM2):
\vspace{-1mm}
\begin{leftbar}
\noindent
\textbf{MM2:} \textit{Frequent and unpredictable 5G-GUTI update.}
\end{leftbar}
\vspace{-1mm}

\subsubsection{5G-S-TMSI-based paging and POs estimate}\label{5G_TMSI_solutions}

Another important improvement of 5G compared to previous generations is the decoupling of the IMSI/SUPI from the paging mechanisms. In 5G, paging takes place with a shortened version of 5G-GUTI, called 5G-S-TMSI (5G S-Temporary Mobile Subscription Identifier) as mentioned in Sec.~2.10.1 of~\cite{Numbering}. 5G-S-TMSI is derived from 5G-GUTI, so its strict update mechanism that was analyzed in the previous section holds for 5G-S-TMSI as well. Based on this outlook, 5G-GUTI-based paging eliminates the IMSI paging attack as analyzed in Sec.~\ref{Sec:IMSI_paging_problem}. Thus, another mitigation mechanism \#3 (MM3) is:
\vspace{-1mm}
\begin{leftbar}
\noindent
\textbf{MM3:} \textit{Decoupling of IMSI from paging.} 
\end{leftbar}
\vspace{-1mm}

Furthermore, the Paging Frame Index (PFI) \textit{is now estimated based on 5G-S-TMSI}, as mentioned in Sec.~7.1 of~\cite{POs_estimate_3GPP} instead of the IMSI. Based on this, the ToRPEDO attack (Sec.~\ref{Sec:ToRPEDO}) is not applicable anymore.
We highlight privacy Mitigation Mechanism \#4 (MM4):
\vspace{-1mm}
\begin{leftbar}
\noindent
\textbf{MM4:} \textit{Decoupling of IMSI from POs' estimate.}
\end{leftbar}
\vspace{-1mm}

\subsection{Improved Integrity \& Confidentiality Protection}\label{Sec:Sol_integrity_confid}

Integrity and confidentiality in 5G networks has many similarities with the status of LTE. First, current 5G Systems are expected to implement the same 128-bit security algorithms with LTE systems: New Radio Encryption Algorithm (NEA) 0, 128-NEA1 and 128-NEA2 for confidentiality (ciphering) and New Radio Integrity Algorithm (NIA) 0, 128-NIA1 and 128-NIA2 for integrity protection, as outlined in Secs.~5.2.2 and~5.2.3 of~\cite{3gpp_fake_base_stations}. On the other hand, the potential use of 256-bit keys has been introduced as an optional feature in 5G~\cite{3gpp_fake_base_stations,256-bit_algorithms_5G_TR} to mitigate potential quantum attacks, such as the Quantum SUCI attack (Sec.~\ref{Sec:suci_quantum}). So, the 5G Mitigation Mechanism \#5 (MM5) is: 
\vspace{-1mm}
\begin{leftbar}
\noindent
\textbf{MM5:}~\textit{256-bit algorithm support for quantum adversary mitigation.}
\end{leftbar}
\vspace{-1mm}

Furthermore, in terms of specific messages protection, it is stated that RRC and NAS integrity are~\textit{mandatory}, whereas NAS and RRC ciphering are~\textit{optional}, something that is similar to LTE. As shown in Sec.~\ref{Sec:PREVIOUS_GENERATIONS}, the attacks described in Secs.~\ref{Sec:RNTI_attacks} and~\ref{Sec:UE_Measurements_attack} can be mitigated by the RRC ciphering so it is added as the Mitigation Mechanism \#6 (MM6) in order to match with these attacks, stating that MM6 is similar in LTE as well:
\vspace{-1mm}
\begin{leftbar}
\noindent
\textbf{MM6 (similar in LTE):}~\textit{Ciphering of RRC messages.} 
\end{leftbar}
\vspace{-1mm}

Moreover, another 5G enhancement compared to LTE is the optional integrity protection of User Plane (UP) messages, as introduced in Sec.~5.3 of~\cite{3gpp_fake_base_stations} that mitigates the Data Manipulation attacks (Sec.~\ref{Sec:Data_Manipulation}). So, the Mitigation Mechanism \#7 (MM7) is:
\vspace{-1mm}
\begin{leftbar}
\noindent
\textbf{MM7:}~\textit{Integrity protection of user plane messages.}
\end{leftbar}
\vspace{-1mm}

\subsection{Privacy of the initial NAS message}\label{Solution_device_capab}
Another important and \textit{mandatory} 5G security mechanism is the privacy of the initial NAS message as described in the section 6.4.6 of~\cite{3gpp_fake_base_stations}. In contrast to previous generations that the whole set of UE Core network capabilities were transmitted plaintext in the initial NAS message, in 5G only the security capabilities are transmitted. The rest of the UE capabilities are transmitted after the activation of a secure NAS channel. This mitigates the Device fingerprinting attack, as described in Sec.~\ref{Device_fingerprinting}. Concluding, the Mitigation Mechanism \#8 (MM8) is:
\vspace{-1mm}
\begin{leftbar}
\noindent
\textbf{MM8}:~\textit{Enhanced privacy for the initial NAS message.}
\end{leftbar}
\vspace{-1mm}

\subsection{Secured Radio Capabilities transmission}\label{Secure_Radio_Capab}

As analyzed in Sec.~\ref{Sec:Radio_Capab_attack}, the UE radio Capabilities were transmitted before the establishment of the RRC security channel, enabling bidding down attacks. This mistaken flow was corrected in 5G, as already mentioned in Sec.~\ref{Sec:Registration_Section}, and depicted in Fig.~\ref{fig:5G_flow}.
This is because the RRC UE Capability Inquiry is transmitted after the RRC Security Mode Complete, when integrity protection is enabled. Based on this outlook, the bidding-down attacks based on UE Radio capabilities are eliminated by the \textit{mandatory} Mitigation Mechanism \#9 (MM9):
\vspace{-1mm}
\begin{leftbar}
\noindent
\textbf{MM9:}~\textit{Radio Capabilities transmission in a secure RRC channel.}
\end{leftbar}
\vspace{-1mm}

\subsection{Fake Base Station Detection Framework}\label{FBS_detection}

Fake (rogue) base stations (FBS) were responsible for many different attacks in the previous cellular generations, as described in Sec.~\ref{Sec:PREVIOUS_GENERATIONS}.
3GPP specifications, for the first time, included an optional detection framework for FBS Detection in the Annex E of~\cite{3gpp_fake_base_stations}. This framework aims to mitigate attacks caused by FBSs, using measurements such as location of cells, frequencies and cell IDs. Although the adversaries are capable of finding the real cell IDs and locations through mobile applications~\cite{alrashede2019imsi}, frequency information can contribute to the detection of traditional active attacks, such as the IMSI catching (Sec.~\ref{Sec:IMSI_Cather}). Furthermore, a whole 3GPP document, TR.~33.809~\cite{TR_FBS_3GPP} is responsible for enhancing the FBS detection framework in future 3GPP releases. For instance, recent 3GPP studies (Secs.~6.4,~6.22 and~6.24~of~\cite{TR_FBS_3GPP}) proposed an enrichment of UE Measurements reports with channel elements, such the RSRP/RSRQ/RSSI of the Channel State Information Reference Signal (CSI-RS) that are capable of detecting and localizing  FBSs. 
In summary, another optional 5G privacy Mitigation Mechanism \#10 (MM10) is:
\vspace{-1mm}
\begin{leftbar}
\noindent
\textbf{MM10:} \textit{Introduction of FBS detection framework.}
\end{leftbar}
\vspace{-1mm}

\section{5G Cellular Generation Attacks}\label{Sec:New_5G_attacks}
In this section, we analyze different attacks against 5G networks based on recent literature. Similarly to Sec.~\ref{Sec:PREVIOUS_GENERATIONS}, we first provide a sub-categorization of different attacks and then we analyze the type of adversaries that are capable of performing these attacks. Finally the protocol vulnerabilities that led to these attacks are outlined. The last seven rows of Table~\ref{Table:Attacks_Overview} indicate our findings.


\subsection{Privacy Attacks}

\subsubsection{\textbf{Identifiers based attacks}}

\paragraph{Quantum SUCI attack:}\label{Sec:suci_quantum}

As explained in the Sec.~\ref{SUCI_mechanism} the SUCI mechanism works with elliptic cryptography. In fact,~\cite{SUCI_QUANTUM} showed that the elliptic cryptography used for SUCI generation can be weak in case of quantum attacks performed by an active adversary, leading to potential SUPI disclosure and identity leakage. This vulnerability has been also discussed by other works~\cite{mitchell2020impact,nguyen2021security,yang2020overview,clancy2019post}, meaning that quantum attacks should be taken into consideration in the next mobile generations. 3GPP has already made some steps, since an optional support of 256-bit key generation algorithms~\cite{3gpp_fake_base_stations} has been considered as a countermeasure to quantum attacks as analyzed in MM6 (Sec.~\ref{Sec:Sol_integrity_confid}).
As shown in~\cite{clancy2019post} a 256-bit encryption algorithm does not mitigate completely potential quantum attacks, but makes their implementation more difficult. Concluding, Weakness \#13 (W13) behind a SUCI Quantum attack is:
\setlength{\leftbarwidth}{4pt}
\setlength{\leftbarsep}{1pt}
\colorlet{leftbarcolor}{red}
\vspace{-1mm}
\begin{leftbar}
\noindent
\textbf{W13:}~\textit{SUCI elliptic cryptography is vulnerable to quantum attacks.} 
\end{leftbar}
\vspace{-1mm}

Furthermore, other attacks against SUCI have been introduced but they do not seem to be a major concern. For example, a SUCI probing or SUCI replay attack~\cite{chlosta20215g} tried to obtain the victim's SUCI in order to verify if a person of interest is in a current location or not. Similarly to IMSI probing attack, 3GPP characterized this attack (Key Issue \#2.2 of~\cite{imsiProbing3GPP}) as low risk and no normative measures are needed. The reason behind this is that obtaining the actual identity of the user cannot be revealed by the SUCI so there is no threat of UE identification. Furthermore, SUPI guessing attacks are described as proof of concepts in~\cite{khan2018identity,liu2021security}.
In this type of attack, they first guess a SUPI and generate SUCIs from this. Then, they send the produced SUCIs to their potential victim, trying to verify if the guessed SUPI belongs to victim user. Such attack has been analyzed as a Key Issue \#3.2 in~\cite{imsiProbing3GPP}, concluding that no normative measures should be taken against this attack since the likelihood of its accuracy is small.
Based on the above, only the SUCI Quantum attack is included in Table~\ref{Table:Attacks_Overview}.

\subsubsection{\textbf{Measurement Exploitation Attacks}}

\paragraph{CSI reports tracking attack}\label{CSI_reports_attack}
An adversary that violates the location privacy of the victim UE based on Channel State Information (CSI) reports has been recently introduced in a 5G network~\cite{ludant2024unprotected}. First, the gnB transmits the Channel State Information Reference Signal (CSI-RS) in the downlink. Then, the UE sends some CSI reports in the Uplink to the gnB as described in the Sec.~5.2 of~\cite{Physical_layer_5G}. The CSI reports include in plaintext information, such as the power of the strongest beam of the CSI-RS which are crucial for the UE positioning. Based on the above, the adversary~\cite{ludant2024unprotected} first constructs a fingerprint database including the power of the strongest beam for the physical area of a gnB. After, the adversary learns the C-RNTI of the victim with the procedures described before (e.g.,~silent messages) and passively decodes the CSI reports including the victim's C-RNTI and strongest beam. Finally, the adversary compares the decoded beam value with the ones existing in the database, thereby estimating the position of the victim and breaking the location privacy. We mention that this attack takes place on the uplink and the existing 5G NR open-source sniffer~\cite{ludant20235g} works only in the downlink. So, commercial and expensive tools should be used for the realization of this attack, such as the WaveJudge passive sniffer~\cite{WaveJudge}. To summarize, the vulnerability that led to this attack is the Weakness \# 14 (W14):
\vspace{-1mm}
\begin{leftbar}
\noindent
\textbf{W14:}~\textit{Lack of encryption of CSI reports.}
\end{leftbar}
\vspace{-1mm}

As stated in~\cite{ludant2024unprotected} CSI reports are an RRC message, so they can take advantage of the potential RRC security context after its activation. On the other hand, the related 3GPP document~\cite{Physical_layer_5G} that refers to the structure of the CSI reports message, does not offer CSI reports encryption. Searching more on the potential encryption of CSI reports we found only a draft of the 3GPP meeting 3GPPSA3\#101-e~\cite{draft_meeting_CSI} that refers to the potential encryption of CSI reports, but nothing exists neither in 3GPP TSs nor 3GPP TRs. So, there is no formal 5G Mitigation Mechanism against the CSI report tracking attack in the 5G standards and 3GPP should consider this in future releases.

\paragraph{ISAC based tracking attack}\label{Sec:ISAC_based_attacks}

Another addition of the Release 19 is the support for Integrated and Sensing (ISAC) applications~\cite{ISAC_3GPP}. ISAC support facilitates emerging 5G and beyond applications, such as health monitoring, smart-homes or environment mapping~\cite{adib2013see,ha2021wistress,gu2018sleepy,vasisht2018duet}. ISAC applications focus on the analysis of wireless signals' reflections (e.g., channel variation in amplitude and phase, signal attenuation, Doppler shift, angle of arrival (AoA)) to infer sensing results. However, a lot of privacy concerns have been raised about ISAC systems in recent literature~\cite{wei2022toward,lu2024integrated,wang2022integrated,liu_sensing,sun2024sok,furqan2021wireless,su2023security}. An active adversary just needs to spoof the legitimate signals to control the sensing applications~\cite{liu2023exploring}, whereas a passive adversary only needs to eavesdrop the wireless channel and sniff the sensing information. As a result, a stealthy passive attacker can estimate the victim's location through existing localization algorithms~\cite{blanco2019performance,vasisht2018duet,kotaru2015spotfi,eleftherakis2024spring+} and potentially learn about the services delivered to the victim (e.g., respiration rate application~\cite{adib2015smart}). A stated in the Sec.~7.1.4 of~\cite{ISAC_3GPP}, 3GPP community has recognized this problem and confidentiality for the sensing data is proposed. On the other hand, aspects on how the implementation of security algorithms will be made, especially if IoT devices are used for sensing are not examined. Furthermore, the 5G Security TS.~\cite{3gpp_fake_base_stations} does not refer to the security of sensing data. As a result, we do not consider any official 5G MM against ISAC based tracking attacks. Thus, Weakness \#15 (W15) is: 
\vspace{-1mm}
\begin{leftbar}
\noindent
~\textbf{W15:}~\textit{Lack of sensing data confidentiality.} 
\end{leftbar}
\vspace{-1mm}

\paragraph{Satellite and NTN tracking attacks:}\label{Sec:NTN_based_attacks}
The integration of satellite and Non Terrestrial networks is considered as an emerging technology for 5G and beyond cellular networks, as stated in relevant literature~\cite{fang20215g,giordani2020non,rinaldi2020non,azari2022evolution,kota20216g,vanelli20205g,kodheli2020satellite} and 3GPP has also performed relevant analysis in several TSs and TRs starting from Release 16~\cite{NTN_SUPPORT_3GPP_TS,NTN_SUPPORT_3GPP_TR,NTN_SUPPORT_3GPP_TR2,NTN_SUPPORT_3GPP_TR_Security}.

However, this promising integration comes with a variety of security and privacy attacks, such as eavesdropping and spoofing~\cite{tedeschi2022satellite,singh2023role,vaezi2022cellular,ahmad2022security,margaria2017signal,chiti2024survey,manulis2021cyber}. 
In fact, satellite communications are usually not security protected, neither in terms of encryption nor in terms of integrity~\cite{pavur2021a}, since the implementation of satellite cryptographic algorithms is quite complex and can reduce rapidly the networks' overall performance~\cite{pavur2020sok,huwyler2023qpep}. 
Thus, the lack of messages encryption is responsible for passive eavesdropping attacks~\cite{pavur2020sok,RECORD_ATTACK}. 
For instance, "RECORD" attack breaks the location privacy of the UE by eavesdropping on the unencrypted wireless signals and messages transmitted by a LEO satellite and correlating them to the user's TMSI~\cite{RECORD_ATTACK}. 
Further, "GSExtract" verified an eavesdropping attack on GEO satellites~\cite{pavur2020tale}. 
So, the weakness behind satellite and NTN based attacks is Weakness \#16 (W16):
\vspace{-1mm}
\begin{leftbar}
\noindent
~\textbf{W16:}~\textit{Lack of confidentiality for satellite and NTN networks data.} 
\end{leftbar}
\vspace{-1mm}

There is a variety of research proposals to improve the security and privacy offered by satellite and NTN networks. 
A first brunch of solutions proposes the application of AI algorithms that aim to detect the existence of an adversary~\cite{iqbal2023empowering,kumar2023distributed,fontanesi2023artificial,mahboob2024revolutionizing,rath2020security,junzhi2019research}. 
Other solutions propose improvements of PHY layer security~\cite{han2022challenges,li2019physical}, such as adding artificial noise to the channel~\cite{formaggio2019authentication,liu2018novel}, and improved cryptographic schemes, such as the adoption of quantum algorithms~\cite{hosseinidehaj2018satellite,bedington2017progress}. 
The above mentioned solutions are not included in the related 3GPP documents~\cite{NTN_SUPPORT_3GPP_TS,NTN_SUPPORT_3GPP_TR,NTN_SUPPORT_3GPP_TR2,NTN_SUPPORT_3GPP_TR_Security}, so no mitigation mechanism exists for these attacks.



\subsection{Security Attacks}
\subsubsection{\textbf{Signal Manipulation attacks}}

\paragraph{PRS spoofing attack:}\label{Sec:PRS_attack}

PRS (Positioning Reference Signal) tracking attacks, that violate the UE location privacy and even hijacks the positioning results of the victim UE is introduced in~\cite{gao2023your,gao2024surgical,focarelli2024physical}. In fact, an active adversary transmits with high power a PRS that interferes with the legitimate PRS signal. As a result, the victim UE considers the hijacked PRS a the legitimate signal, since there is no integrity mechanism in the PHY layer to show the PRS modification. Then, this hijacked PRS is used by the 5G system for the UE localization. So, the adversary not only violates the UE location privacy, but also controls the position of the victim. This attack is caused by  Weakness \#17 (W17):
\vspace{-1mm}
\begin{leftbar}
\noindent
\textbf{W17:}~\textit{Lack of integrity protection in the PHY layer.} 
\end{leftbar}
\vspace{-1mm}

We mention that even after the activation of the security context, physical signals cannot be easily integrity protected due to high computational overhead and large number of protocol modifications. The solution for this attack is a strong FBS detection framework and for this reason the same paper~\cite{gao2023your} proposes an enhanced version of the FBS detection framework, that uses Machine Learning and detects the PRS tracking attack with high accuracy.

\paragraph{Ambient-IoT device spoofing attack:}\label{Sec:AmbientIoT_attack}
Technically moving beyond typical UEs, the latest Release 19~\cite{ambient_IoT_3GPP,ambient_IoT_Technical_Report} of 5G-Advanced has introduced the implementation of ambient-IoT devices.
Ambient-IoT devices, such as healthcare sensors, are battery-free devices that essentially operate using environmental energy through backscatter method~\cite{varshney2017lorea,talla2017lora,liu2013ambient}. A recent 3GPP meeting~\cite{ambient_IoT_meeting} referred to the power consumption of ambient IoT devices, stating that ambient IoT devices of category $1$ have a consumption of $1$ $\mu$W. Due to their limited energy, security algorithms are difficult to be implemented in ambient-IoT devices, raising privacy and security concerns~\cite{van2018ambient_survey,jiang2023backscatter}.
Interestingly, an active attacker can spoof the backscatter signals controlling the ambient-IoT device, thereby breaking both the location and untraceability properties. Initial attempts to detect such attacks relied on ambient IoT device fingerprinting, based on hardware characteristics such as the radio frequency offsets~\cite{zanetti2010physical}, which as shown in ~\cite{narayanan2021harvestprint} are not accurate enough due to channel instability.
On the other hand, HarvestPrint system~\cite{narayanan2021harvestprint} fingerprints the ambient IoT devices based on their energy discharge, which is differentiated compared to the random radio frequency offsets, thus detecting spoofing attacks accurately. Since Section 5.2.6 of~\cite{ambient_IoT_3GPP}, which analyzes security and privacy aspects of ambient IoT devices, is in a preliminary stage, it is unclear how 3GPP plans to mitigate the existing privacy problems. 
Therefore, there is no official 5G MM for this attack. Concluding, the Weakness that causes this attack is Weakness \#18 (W18): 
\vspace{-1mm}
\begin{leftbar}
\noindent
~\textbf{W18:}~\textit{Lack of security algorithms for ambient-IoT devices.} 
\end{leftbar}
\vspace{-1mm}

\subsubsection{\textbf{Protocol Downgrade attacks}}

\paragraph{5G Bidding Down attack:}\label{Sec:5G_bidding_down_attack}
A 5G bidding down attack was introduced, as a proof of concept in~\cite{hussain20195greasoner}. An active adversary using a FBS eavesdrops the RRC Security Mode Command sent by the legitimate gNB to the UE and then answers on behalf of the victim UE with a RRC Security Mode Failure. As stated in the Sec.~5.3.4.3 of~\cite{3GPP_RRC}, after an RRC Security Mode Failure message reception, the security algorithms used prior to the reception of the RRC Security Mode Command message continue to be used. As a result, no integrity and confidentiality are applied and the UE works in limited mode with "NULL" ciphering and integrity algorithms (NEA0 and NIA0). Finally after deactivating the victim's UE security protection, the adversary may send an identity request message to the victim and steal its SUPI-IMSI, thus breaking the identity privacy as well.
Concluding, the Weakness \#19 (W19) behind this attack is: 
\vspace{-1mm}
\begin{leftbar}
\noindent
\textbf{W19:}~\textit{Lack of integrity protection in RRC security mode failure message}. 
\end{leftbar}
\vspace{-1mm}

Considering that this attack works before the activation of an RRC secure channel, integrity protection cannot be activated for this message. This attack among with the one described in Sec.~\ref{Sec:PRS_attack} highlight the need for a strong and compulsory FBS detection framework (MM10). For example, a recent work~\cite{scalingi2024det} proposes a RAN based FBS detection framework, that is capable of mitigating bidding down attacks with the application of Deep Learning techniques. 

\section{Discussion}\label{Sec: Discussion}

In this paragraph, we comment on the results presented in the previous Sections~4, 5 and 6, on where 5G security stands compared to its predecessors, and extract important takeaways.

5G standard paid particular attention to 5G-GUTI mechanism and paging procedures. As a result, IMSI has been substituted by 5G-GUTI for paging and POs' estimate as explained in Sec.~\ref{5G_TMSI_solutions}. These~\textit{mandatory} modifications along with the strict and also~\textit{mandatory} 5G-GUTI reallocation mechanism have mitigated some common pre-5G attacks (Secs.~\ref{Sec:IMSI_paging_problem},~\ref{Sec:ToRPEDO},~\ref{Sec:guti_problems}).

\setlength{\leftbarwidth}{5pt}
\setlength{\leftbarsep}{8pt}
\colorlet{leftbarcolor}{teal}
\vspace{-1mm}
\begin{leftbar}
\noindent
\textbf{Takeaway 1:}
5G ensures a secure paging procedure through the mandatory 5G-GUTI mechanism.
\end{leftbar}

5G focused on the privacy of the UE CN and Radio Capabilities.
In fact, the enhanced privacy for the first NAS message (Sec.~\ref{Solution_device_capab}) and the transmission of Radio Capabilities after the establishment of a secure channel (Sec.~\ref{Secure_Radio_Capab}) mitigate attacks related to UE CN and Radio Capabilities as described in Sec.~\ref{Sec:Radio_Capab_attack} and~\ref{Device_fingerprinting}.
\vspace{-1mm}
\begin{leftbar}
\noindent\textbf{Takeaway 2:}
5G enhances the privacy of UE Core Network and Radio Capabilities.
\end{leftbar}

SUCI (Sec.~\ref{SUCI_mechanism}) was introduced as an optional 5G security mechanism that is capable of mitigating the IMSI catching (Sec.~\ref{Sec:IMSI_Cather}) and IMSI extractor (Sec.~\ref{Sec:IMSI_Extractor}) attacks.
However, SUCI has been shown to be vulnerable to quantum adversaries (Sec.~\ref{Sec:suci_quantum}), and the use of quantum-resistant encryption algorithms is optional as well (MM5 in Sec.~\ref{Sec:Sol_integrity_confid}). 
\vspace{-1mm}
\begin{leftbar}
\noindent \textbf{Takeaway 3:}
The SUCI mechanism, key for IMSI/SUPI confidentiality, is still optional and vulnerable to quantum attacks. 3GPP should consider enforcing both SUCI and quantum resistant cryptography algorithms to ensure user identity privacy.
\end{leftbar}

5G offers optional integrity of UP data (MM8) against data manipulation attacks (Sec.~\ref{Sec:Data_Manipulation}).
A recent work~\cite{heijligenberg2023bigmac} showed that the privacy enhancement offered by UP integrity comes with the increase of the network computational overhead and latency. Parallel programming, more computationally efficient integrity algorithms and CN architecture improvements are proposed to minimize the networks' overhead~\cite{heijligenberg2023bigmac}. 
\vspace{-1mm}
\begin{leftbar}
\noindent \textbf{Takeaway 4:} UP integrity is another new but optional 5G MM. 3GPP should consider potential ways to minimize the computational overhead added by UP integrity and then mandate it.
\end{leftbar}

As shown in Sec.~\ref{Sec:Sol_integrity_confid}, the ciphering of RRC messages (MM7) is optional in 5G as it was in LTE. As a result, the mitigation of C-RNTI catching (Sec.~\ref{Sec:RNTI_attacks}) and UE measurement reports attacks (Sec.~\ref{Sec:UE_Measurements_attack}) is operators' choice. To make matters worse, an open-source 5G sniffer is available~\cite{ludant20235g} and has already performed the C-RNTI tracking attack in 5G networks with high accuracy. Another solution, except of mandating RRC ciphering, is the most frequent reallocation of C-RNTI, through a pool of multiple C-RNTIs as proposed in recent literature~\cite{Variable_RNTI_solution,ludant20235g,saeed2021identity}. All~\cite{Variable_RNTI_solution,ludant20235g,saeed2021identity} can be 3GPP compliant with minimum modifications and their computational overhead is lower than mandating RRC ciphering.
\vspace{-1mm}
\begin{leftbar}
\noindent \textbf{Takeaway 5:} 5G standard offers no improvement in terms of RRC ciphering compared to LTE, thus being vulnerable to the same inherited attacks.
\end{leftbar}

Based on our findings, the only inherited attack that is not defended by even an optional 5G MM is the AKA Linkability attack (Sec.~\ref{Sec:AKA_Linkability_Attack}).
As a result, even the new 5G-AKA protocol is vulnerable to this attack~\cite{fei2023vulnerability,wang2021privacy,hussain20195greasoner,koutsos20195g}.
In fact, 3GPP community analyzed this attack in the Sec.~6.2 of~\cite{suci_related_solutions}; thus, the ciphering of both failure messages and different 5G-AKA challenges (e.g.,~AUTN) was proposed. These modifications can successfully mitigate the problem of AKA Linkability Attack, but they are still not included in the related 3GPP TS.~\cite{3gpp_fake_base_stations}. 
\vspace{-1mm}
\begin{leftbar}
\noindent \textbf{Takeaway 6:}
AKA linkability attack has not been mitigated in 5G and future releases should consider this problem.
\end{leftbar}

A new 5G specific location tracking attack has been shown recently, based on CSI reports (Sec.~\ref{CSI_reports_attack}).
As explained earlier and illustrated in Table~\ref{Table:Attacks_Overview}, no 5G MM exists for defending this attack.
\vspace{-1mm}
\begin{leftbar} 
\noindent\textbf{Takeaway 7:}
3GPP should evaluate the CSI reports attack and propose corresponding mitigation mechanisms in future releases for encrypting CSI reports.
\end{leftbar}

Recent 3GPP Releases introduced ISAC-based applications, satellite/NTN networks' integration and Ambient-IoT. As explained in Secs. ~\ref{Sec:ISAC_based_attacks},~\ref{Sec:NTN_based_attacks} and~\ref{Sec:AmbientIoT_attack} and illustrated in Table~\ref{Table:Attacks_Overview}, these applications encounter serious privacy and security risks and no MMs exist for users' protection.
\vspace{-1mm}
\begin{leftbar} 
\noindent\textbf{Takeaway 8:}
3GPP should propose security and privacy frameworks for emerging technologies, such as Ambient-IoT and ISAC-based applications and satellite/NTN networks integration, in future releases.
\end{leftbar}

FBS detection framework is the only 5G MM that is capable of mitigating stealthy attacks, such as the ones presented in Secs. \ref{Sec:IMSI_Extractor},~\ref{Sec:PRS_attack} and~\ref{Sec:5G_bidding_down_attack}.
As explained in Sec.~\ref{FBS_detection}, 5G introduced an optional FBS detection framework and recent studies enrich it~\cite{TR_FBS_3GPP}. Besides, there are many proposals for the enhancement of the FBS detection framework in recent literature that should be taken into consideration. A first set of solutions~\cite{lotto2023baron,hussain2019insecure,singla2021look} propose the authentication of legitimate BSs by defining new authentication protocols, network entities or digital signature schemes, thereby mitigating the impact of FBS but increasing both the computational overhead of the network and soliciting protocol modifications.
On the other hand, many recent works focus on the application of Artificial Intelligence (AI) for FBS detection ~\cite{li2021secure,scalingi2024det,gao2023your,nakarmi2021murat,mubasshir2024fbsdetector,nakarmi2022applying,saedirbs} with remarkable results.
In particular, the AI-based FBS frameworks proposed by ~\cite{gao2023your,scalingi2024det} are capable of detecting even new and sophisticated attacks, as analyzed in Sec.~\ref{Sec:PRS_attack} and~\ref{Sec:5G_bidding_down_attack}. The Open RAN (O-RAN) evolution can facilitate the implementation of these AI solutions in the current 5G networks~\cite{polese2023understanding}.
\vspace{-1mm}
\begin{leftbar}    
\noindent \textbf{Takeaway 9:}
FBS detection framework is necessary for enhancing UE privacy. 3GPP should define a strict and mandatory framework in future releases, utilizing AI advancements.
\end{leftbar}   

Finally, we compare different network infrastructures based on recent measurements studies.
Current cellular networks operate either in 5G Stand Alone (SA) or 5G NoN-Stand Alone (NSA). Recent 5G NSA measurement studies~\cite{lasierra2023european,park20215g,park2022analyzing,karakoc2023never,cheng2023watching,kwon2021towards,wani2024security} made in real networks show that 5G NSA poses significant privacy risks to the UE privacy. This can be explained by the fact that 5G NSA networks follow an LTE CN. 
Thus, only 5G enhancements related with the RAN~\cite{Lenders1,Lenders2} can be integrated.
In contrast, the only available 5G SA measurement study~\cite{nie2022measuring} showed significant improvements, since mandatory 5G MMs are used and ciphering/integrity protection is activated to a higher rate. 
Unfortunately, incompatibility issues between the 5G CN and the USIM cards~\cite{yu2023secchecker} or USIM cards miss-functionalities~\cite{bitsikas2023ue} limit the SUCI implementation~\cite{nie2022measuring}.
\vspace{-1mm}
\begin{leftbar}
\noindent \textbf{Takeaway 10:}
Operators should update their networks from 5G NSA to 5G SA: 5G NSA cannot follow the majority of 5G MMs. USIM editors need to correct incompatibility issues with SUCI mechanism.
\end{leftbar}
\section{Conclusion}

In this systematization of knowledge paper, we first introduced a methodology for studying privacy attacks and corresponding mitigation mechanisms (MMs) in cellular networks.
We studied 12 pre-5G attacks and 10 5G MMs that were proposed as a solution by 5G Standards and Protocols.
As we remark, 1 of these attacks has been resolved since LTE, and 5 are resolved by mandatory 5G MMs.
On the other hand, 5 attacks are mitigated by optional MMs, thereby making their implementation an operator's choice.
Furthermore, the AKA Linkability attack continues to persist even in 5G, since it is not resolved by any 5G MM.
Moreover, we studied 6 novel and recent 5G attacks, 3 of which cannot be resolved by the current 5G protocols, whereas the rest can be resolved only by optional MMs. Finally, we outlined 10 key takeaways from our analysis, including some lessons learned from our study and making proposals for future work both for operators and the 3GPP community.

\section*{ACKNOWLEDGEMENTS}
This research was partially supported by:
The Ministry of Economic Affairs and Digital Transformation of Spain and the European Union-Next Generation EU programme for the Recovery, Transformation and Resilience Plan and the Recovery and Resilience Mechanism under agreements TSI-063000-2021-63 (MAP-6G), TSI-063000-2021-142 and TSI-063000-2021-147 (6G-RIEMANN);
The European Union Horizon 2020 program under grant agreements 101021808 (SPATIAL), 101096435 (CONFIDENTIAL6G) and 101139067 (ELASTIC).
The views and opinions expressed are those of the authors only and do not necessarily reflect those of the European Union. Neither the European Union nor the granting authority can be held responsible for them.

\newpage
\bibliographystyle{ieeetr}
\bibliography{referencesSOK}

\end{document}